\documentclass[final,12pt]{elsarticle}

\usepackage{geometry}
\usepackage{graphicx}

\makeatletter
\def\ps@pprintTitle{%
 \let\@oddhead\@empty
 \let\@evenhead\@empty
 \let\@oddfoot\@empty
 \let\@evenfoot\@empty}
\makeatother

\usepackage{xcolor,colortbl}

\usepackage{tikz}
\usepackage{array}

\usepackage{longtable}
\usepackage{pifont}
\newcommand{\xmark}{\ding{55}}  

\usepackage{booktabs}

\usepackage{tabularx} 

\usepackage{amssymb}

\journal{Computers \& Security}

\begin{document}

\begin{frontmatter}

\title{Safeguarding Connected Autonomous Vehicle Communication: Protocols, Intra- and Inter-Vehicular Attacks and Defenses}

\author[inst1]{Mohammed Aledhari}
\author[inst2]{Rehma Razzak}
\author[inst3]{Mohamed Rahouti}
\author[inst8]{Abbas Yazdinejad}
\author[inst9]{Reza M.Parizi}
\author[inst4]{Basheer Qolomany}
\author[inst5]{Mohsen Guizani}
\author[inst6]{Junaid Qadir}
\author[inst7]{Ala Al-Fuqaha}

\affiliation[inst1]{organization={Department of Data Science, University of North Texas},
            addressline={Denton}, 
            state={TX},
            postcode={76207},             
            country={USA,},
            email={ mohammed.aledhari@unt.edu}}

\affiliation[inst2]{organization={College of Computing and Software Engineering, Kennesaw State University},
            addressline={Marietta}, 
            state={GA},
            postcode={30060},             
            country={USA,},
            email={ rrazzak@students.kennesaw.edu}}

\affiliation[inst3]{organization={Department of Computer and Information Science, Fordham University},
            addressline={New York}, 
            state={NY},
            postcode={10023},             
            country={USA,},
            email={ mrahouti@fordham.edu}}

\affiliation[inst8]{organization={Cyber Science Lab, Canada Cyber Foundry, University of Guelph},
              postcode={N1G4S7}, 
            state={ON},
            country={Canada,},
            email={ ayazdine@uoguelph.ca}}
            
\affiliation[inst9]{organization={Decentralized Science Lab, Kennesaw State University},
            state={GA},
            country={USA,},
            email={ rparizi1@kennesaw.edu}}

\affiliation[inst4]{organization={Department of Medicine, College of Medicine, Howard University},
            addressline={Washington}, 
            state={DC},
            postcode={20059}, 
            country={USA,},
            email={ basheer.qolomany@howard.edu}}

\affiliation[inst5]{organization={Machine Learning Department, Mohamed Bin Zayed University of Artificial Intelligence (MBZUAI)},
            addressline={Abu Dhabi},
            country={UAE,},
            email={ mguizani@ieee.org}}
            
\affiliation[inst6]{organization={Computer Science and Engineering Department, Qatar University},
            addressline={Doha},
            country={Qatar,},
            email={ jqadir@qu.edu.qa}}

\affiliation[inst7]{organization={Information and Computing Technologies (ICT) Division, College of Science and Engineering (CSE), Hamad Bin Khalifa University},
            addressline={Doha},
            country={Qatar,},
            email={aalfuqaha@hbku.edu.qa}}
\begin{abstract}

\textcolor{black}{The advancements in autonomous driving technology, coupled with the growing interest from automotive manufacturers and tech companies, suggest a rising adoption of Connected Autonomous Vehicles (CAVs) in the near future. Despite some evidence of higher accident rates in AVs, these incidents tend to result in less severe injuries compared to traditional vehicles due to cooperative safety measures. However, the increased complexity of CAV systems exposes them to significant security vulnerabilities, potentially compromising their performance and communication integrity. This paper contributes by presenting a detailed analysis of existing security frameworks and protocols, focusing on intra- and inter-vehicle communications. We systematically evaluate the effectiveness of these frameworks in addressing known vulnerabilities and propose a set of best practices for enhancing CAV communication security. The paper also provides a comprehensive taxonomy of attack vectors in CAV ecosystems and suggests future research directions for designing more robust security mechanisms. Our key contributions include the development of a new classification system for CAV security threats, the proposal of practical security protocols, and the introduction of use cases that demonstrate how these protocols can be integrated into real-world CAV applications. These insights are crucial for advancing secure CAV adoption and ensuring the safe integration of autonomous vehicles into intelligent transportation systems.} 
\end{abstract}
\end{frontmatter}

\section{Introduction} \label{intro} 
\begin{figure*} [htbp]
\centering
\includegraphics [width=.95\textwidth] {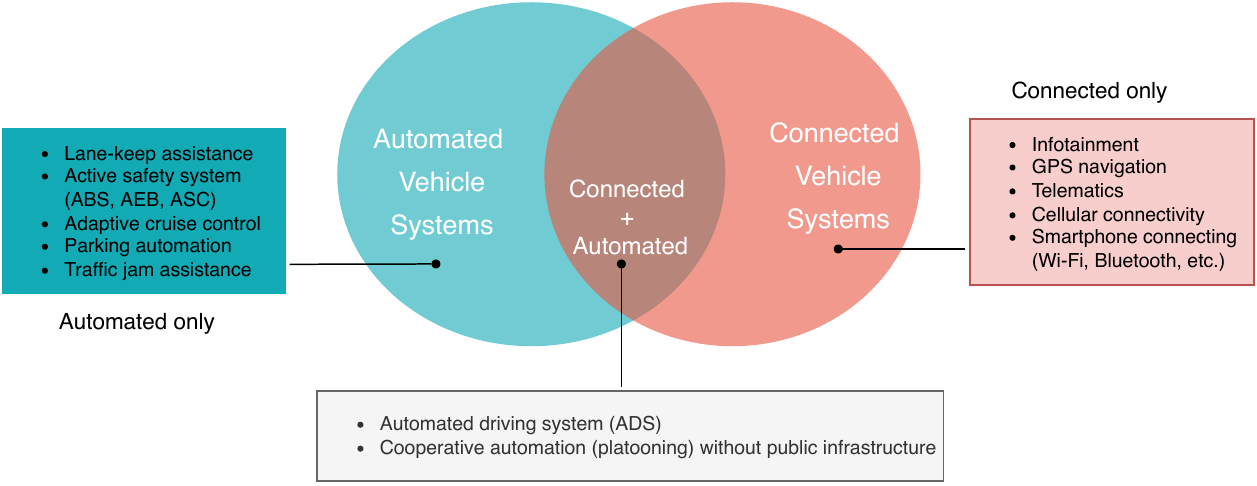}
\caption {Connected Autonomous Vehicle (CAV) infrastructure.}
\label{CAVconcepts}
\end{figure*}

Connected Autonomous Vehicles (CAVs) are a progression of various forms of autonomous technology \cite{fagnant2015preparing}. \textcolor{black}{CAVs enable vehicles to navigate from source to destination with limited or no human involvement, providing significant advancements in automation and connectivity.} Self-driving cars can be categorized into several levels of autonomy, ranging from 0 to 5 \cite{litman2017autonomous}. At each level, the CAV gains more autonomous properties, with Level 0 being fully human-controlled, and Level 5 representing full automation without human input \cite{narbayeva2020blockchain}, \cite{kalra2016driving}. \textcolor{black}{While connectivity and automation are distinct aspects of CAVs, they often overlap, leading to complex communication requirements between vehicles and infrastructure.} Figure \ref{CAVconcepts} illustrates these differences and overlaps. CAVs promise numerous benefits, notably enhancing transportation safety and efficiency \cite{hodge2019vehicle}. However, they also introduce significant security challenges, particularly in managing communication and control systems that involve sensing, computing, and vehicular inter-/intra-communication \cite{calandriello2010performance}. 

\textcolor{black}{CAVs have been the target of various cyberattacks, particularly against critical vehicular systems such as sensors, GPS, and control units.} Over 126 cyber incidents have been reported against autonomous vehicles \cite{chowdhury2020attacks}. For example, the Nissan Leaf's mobile app vulnerability allowed hackers to access its heater and battery systems, leading to the app's shutdown \cite{al2019autonomous}. Similarly, hackers have exploited vulnerabilities in BMW and Tesla vehicles, causing malfunctions and safety risks \cite{chowdhury2020attacks}. These attacks highlight the growing threat to CAV communication systems, especially those responsible for self-driving decision-making.

\begin{figure*}[h]
    \centering
	\includegraphics[width=6in]{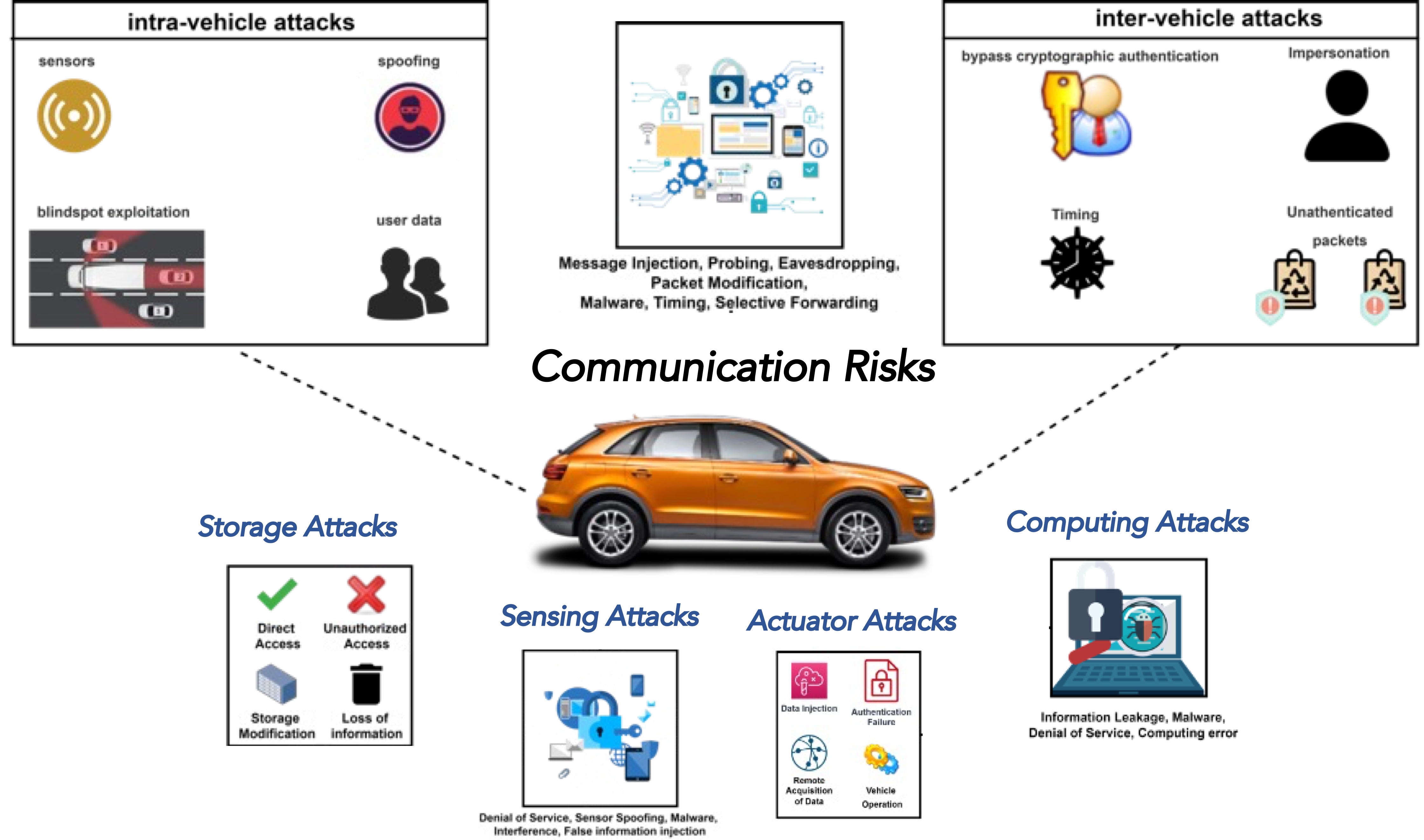}
	\caption{Attack surface in CAVs. \textit{Top: communication risks (intra-vehicular and inter-vehicular attack)}.}
	\label{fig:attacks}
\end{figure*}

Securing CAVs remains a critical and open challenge \cite{abraham2016autonomous}. \textcolor{black}{The complexity of CAVs, combined with their reliance on multiple interconnected systems, increases their vulnerability to cyberattacks. Ensuring the secure operation of these vehicles requires robust legal and technical frameworks.} Currently, there are no specific regulations mandating cybersecurity for CAVs \cite{cite-regulations}. \textcolor{black}{As the amount of software in vehicles grows, so too does the potential attack surface, making it essential to implement advanced security measures throughout the design and operation of CAVs.} Hackers can disrupt safety-critical systems, leading to severe consequences, such as loss of control and traffic accidents \cite{cite-hackers}. Figure \ref{fig:attacks} provides an overview of potential attack surfaces in CAVs.

\textcolor{black}{CAV communication faces numerous threats, including attacks on Vehicle-to-Vehicle (V2V) and Intra-vehicle communication systems.} Inter-vehicle communication shares data about traffic, accidents, and road conditions \cite{tsugawa2002inter}, while Intra-vehicle communication relays information between sensors and control units \cite{el2020cybersecurity}. Both systems are vulnerable to eavesdropping, spoofing, man-in-the-middle attacks, and more. Recent incidents involving BMW and Tesla vehicles have exposed these vulnerabilities, leading to unauthorized control and unpredictable behaviors such as sudden braking or lane changes \cite{noauthor_upstream_2020}.

These examples prove how crucial it is to distinguish between what type of attack has occurred to mitigate it and properly prevent its future occurrences \cite{viriyasitavat2015vehicular, singh2015secure, ying2015motivation}. Identifying certain types of attacks can be difficult as they may target the overall infrastructure of the CAV \cite{abbott2016techniques}. Classification of attacks on CAVs is of critical importance as each attack can be associated with its own working mechanisms \cite{hussain2018autonomous}, \cite{koopman2016challenges}. To rectify this, we would first need to identify the most common attack vectors. Attack vectors are means by which hackers gain access, and a single incident can include multiple attack vectors. A report done by Upstream revealed that the most common attack vectors for CAVs included servers, keyless entry systems, and mobile apps. Additionally, the OBD (On-board Diagnostics) port, IT networks, sensors, and infotainment systems were also among the most common attack vectors for CAVs. Another unsettling discovery from the Upstream Report was that at least 80\% of the attacks against CAVs were remote. This concern has still not been properly addressed because remote attacks against CAVs are harder to prevent and can lead to significantly more damage. Another step would be to categorize the attacks on CAVs in severity. A categorization of these attacks comes from the  ISO/SAE 21434 standard, which requires vehicle manufacturers and suppliers to ensure that cybersecurity risks are managed efficiently and effectively. This standard contains the Threat Analysis and Risk Assessment (TARA), which determines the extent to which a road user can be impacted by automotive cyber threats and vulnerabilities. The ISO/SAE 21434 standard also explains that one way to determine this is to use the CVSS 3.X (Common Vulnerability Scoring System) exploit-ability score, which has values that range between 0.12 (very low) and 3.89 (high).

\textcolor{black}{Furthermore, traditional security solutions cannot be directly applied to CAV communication due to the following reasons:
\begin{itemize}
    \item The number of software gradually increases in CAVs, which requires higher levels of security built into the vehicular technology from the ground up.
    \item CAVs share and exchange a high volume of data and decision-making-specific messages by accessing a digital infrastructure involving access to the public Internet. This information sharing, including personal data, can potentially broaden the range of security vulnerabilities in CAVs' communications.
    \item The ability to use traditional encryption and authentication solutions (from other industries) in CAVs is restricted as they impose unique requirements and limitations.
    \item Traditional intrusion detection systems (IDSs) typically deploy static analysis approaches, which could be inefficient in defending against the deformation of security attacks when exposing security loopholes.
    \item The enabling of third-party apps in CAVs is continuously increasing, making it challenging to ensure communication security due to the absence of security by design.
    \item Traditional Machine Learning (ML)--based attack detection solutions typically rely on conventional feedback controllers and can potentially degrade driving safety by generating inaccurate decisions over the varied road conditions and traffic dynamics.
\end{itemize}}

To alleviate potential security threats, countermeasures must be implemented effectively for CAVs \cite{maimaris2016review}. Unfortunately, adopting them can be ineffective or misguided for a few reasons \cite{wang2018networking}:
\begin{enumerate}
\item Depending on existing security guidelines and the assumed adversarial model(s), certain weaknesses may be unnecessary to consider.
\item When generalizing attacks across all CAVs, there is a tendency to ignore security perimeter modeling (SPM), a fundamental principle for designing security architectures for CAVs.
\end{enumerate}

Beyond attacks, current networking and security protocols, techniques for securing communication, and other possible factors must be examined. Many security frameworks have been designed and implemented by researchers, from which best practices for designing effective security frameworks to achieve maximum security without sacrificing computational resources or user comfort can be generalized. A few use cases have also been developed to tackle some security issues relating to CAVs.

\subsection{Scope of the Survey}
This paper focuses on intra- and inter-vehicle communication security, discussing the state-of-the-art security frameworks, challenges, and techniques associated with CAV communication. Particular attention is given to the architectures, protocols, and frameworks that secure these systems at both levels. \textcolor{black}{Given the substantial progress in CAV security, this article provides a comprehensive review of current technologies and explores new avenues for enhancing communication security in autonomous driving systems.} \textcolor{black}{By highlighting technical robustness and assurance requirements, this survey offers a detailed tutorial on the latest developments in securing CAV communications.} \textcolor{black}{The analysis presented here provides researchers and engineers with an up-to-date review of the most pressing security issues in CAV communication, while offering insights into potential solutions and unexplored research areas.}

\subsection{Literature Selection Methodology}
\textcolor{black}{To ensure a comprehensive and up-to-date review of the literature, a systematic methodology was employed in selecting relevant papers. The primary databases used for literature retrieval were IEEE Xplore, Google Scholar, and major academic journals in automotive cybersecurity. We prioritized papers published after 2015 to capture the latest advancements in CAV security, while also including seminal works that have laid the foundation for current research in this domain. The selection process emphasized papers relevant to intra- and inter-vehicular communication security, ensuring that each study offered significant insights into the challenges and solutions within this field.} \textcolor{black}{Our inclusion criteria also focused on papers that contribute to the development of new protocols, attack detection methods, and security enhancements for CAV systems. We ensured a balanced review by covering both cutting-edge technologies, such as machine learning-based security solutions, and traditional approaches like cryptographic protocols. Additionally, the relevance and impact of each paper were evaluated based on the proposed security frameworks and their applicability to real-world CAV communication challenges.}

\subsection{Related Papers}
\textcolor{black}{In this section, we review the existing literature on communication security in CAVs. Numerous studies have been conducted on different aspects of CAV security, ranging from attack taxonomies to proposed security frameworks and protocols. The reviewed literature spans both theoretical and empirical studies, with a focus on contributions that propose novel security frameworks, address communication vulnerabilities, or provide comprehensive surveys of the field. Key areas explored in the literature include cryptographic protocols, secure communication architectures, machine learning-based threat detection, and privacy-preserving mechanisms for CAVs. Notable works include both recent advancements published after 2015 and seminal papers that have significantly influenced the field of CAV security.}

\tiny
\begin{longtable}{
    |>{\centering\arraybackslash}p{1.8cm}
    |>{\centering\arraybackslash}p{0.7cm}
    |>{\centering\arraybackslash}p{0.6cm}
    |>{\centering\arraybackslash\arraybackslash}p{0.6cm}
    |>{\centering\arraybackslash}p{0.6cm}
    |>{\centering\arraybackslash}p{0.6cm}
    |>{\centering\arraybackslash}p{0.6cm}
    |>{\centering\arraybackslash}p{0.8cm}
    |>{\centering\arraybackslash}p{0.7cm}
    |>{\centering\arraybackslash}p{0.6cm}
    |>{\centering\arraybackslash}p{5.6cm}
    |}
\caption{Summary of existing surveys and reviews related to CAV Security.} \label{tab:survey} \\
\hline
\textbf{Refs. (Author)} & \rotatebox{90}{\textbf{Year of Pub.}} & \rotatebox{90}{\textbf{CAV Review}} & \rotatebox{90}{\textbf{Sec. \& Priv.}} & \rotatebox{90}{\textbf{Attack Vect.}} & \rotatebox{90}{\textbf{Inter-Comm.}} & \rotatebox{90}{\textbf{Intra-Comm.}} & \rotatebox{90}{\textbf{Auton. Driving Apps}} & \rotatebox{90}{\textbf{Eval. Tools}} & \rotatebox{90}{\textbf{Standards}} & \textbf{Overview} \\ \hline
\endfirsthead

\multicolumn{11}{c}%
{{\bfseries \tablename\ \thetable{} -- continued from previous page}} \\
\hline
\textbf{Refs. (Author)} & \rotatebox{90}{\textbf{Year of Pub.}} & \rotatebox{90}{\textbf{CAV Review}} & \rotatebox{90}{\textbf{Sec. \& Priv.}} & \rotatebox{90}{\textbf{Attack Vect.}} & \rotatebox{90}{\textbf{Inter-Comm.}} & \rotatebox{90}{\textbf{Intra-Comm.}} & \rotatebox{90}{\textbf{Auton. Driving Apps}} & \rotatebox{90}{\textbf{Eval. Tools}} & \rotatebox{90}{\textbf{Standards}} & \textbf{Overview} \\ \hline
\endhead

\hline \multicolumn{11}{|r|}{{Continued on next page}} \\ \hline
\endfoot

\hline \hline
\endlastfoot


Papadimitratos \textit{et al.} \cite{papadimitratos2008secure} & 2008 & \xmark{} & \checkmark{} & \checkmark{} & \xmark{} & \xmark{} & \xmark{} &  \xmark{} & \xmark{} & Reviews threats and types of adversaries, privacy and security requirements, and mechanisms to secure vehicular systems \\ \hline


Miller \textit{et al.} \cite{miller2014survey} & 2014 & \xmark{} & \xmark{} & \xmark{} & \checkmark{} & \checkmark{} & \xmark{} &  \xmark{} & \checkmark{} & Features automotive remote attacks and effects of these attacks on automotive cyber-physical surfaces including defensive and precautions to administer \\ \hline


Viriyasitavat \textit{et al.} \cite{viriyasitavat2015vehicular} & 2015 & \xmark{} & \xmark{} & \xmark{} & \checkmark{} & \checkmark{} & \checkmark{} &  \xmark{} & \checkmark{} & Provides vehicular communication and implementation in dynamic environments with associated propagation and channel modeling challenges  \\ \hline


Gupta \textit{et al.} \cite{gupta2015survey} & 2015 & \checkmark{} & \xmark{} & \xmark{} & \checkmark{} & \checkmark{} & \checkmark{} &  \xmark{} & \xmark{} & Highlights mobile communication networks and their applications to civil and military domains \\ \hline


Abraham \textit{et al.} \cite{abraham2016autonomous} & 2016 & \xmark{} & \xmark{} & \xmark{} & \checkmark{} & \checkmark{} & \xmark{} &  \checkmark{} & \xmark{} & Surveys AV consumer preferences particularly mobility and reduction of accidents \\ \hline


Thing \textit{et al.} \cite{thing2016autonomous} & 2016 & \xmark{} & \xmark{} & \checkmark{} & \xmark{} & \checkmark{} & \xmark{} &  \xmark{} & \xmark{} & Presents security attack mechanisms against AV along with a taxonomy of their corresponding defense strategies \\ \hline


Gifei \textit{et al.} \cite{gifei2017integrated} & 2017 & \checkmark{} & \checkmark{} & \xmark{} & \xmark{} & \xmark{} & \xmark{} &  \checkmark{} & \xmark{} & Reviews ECUs and the integration of management systems for quality, safety and security purposes in the development of AVs \\ \hline


C{\u{a}}ilean \textit{et al.} \cite{cuailean2017current} & 2017 & \xmark{} & \xmark{} & \xmark{} & \checkmark{} & \checkmark{} & \checkmark{} &  \xmark{} & \checkmark{} & Emphasizes challenges for visible light communications in vehicle applications with solutions and supportive data \\ \hline


Karnouskos \textit{et al.} \cite{karnouskos2017privacy} & 2017 & \xmark{} & \checkmark{} & \xmark{} & \checkmark{} & \xmark{} & \xmark{} &  \xmark{} & \xmark{} & Discusses the feasibility of ensuring integrity, safety, and privacy in hyper-connected vehicle scenarios\\ \hline


Parkinson \textit{et al.} \cite{parkinson2017cyber} & 2017 & \xmark{} & \xmark{} & \checkmark{} & \xmark{} & \xmark{} & \checkmark{} &  \xmark{} & \checkmark{} & Reviews a large volume of accessible literature and compartmentalise existing security challenges based on the state-of-art vulnerabilities countermeasure techniques \\ \hline


Claybrook \textit{et al.} \cite{claybrook2018autonomous} & 2018 & \checkmark{} & \xmark{} & \xmark{} & \xmark{} & \xmark{} & \xmark{} &  \xmark{} & \checkmark{} & Provides statistical data regarding number of vehicle accidents and the safety benefits of AVs to due to the implemented federal regulations \\ \hline


De Bruyne \textit{et al.} \cite{de2018merging} & 2018 & \checkmark{} & \xmark{} & \xmark{} & \xmark{} & \xmark{} & \xmark{} &  \xmark{} & \checkmark{} & Introduces regulation policies to self-driving cars by analyzing the legal issues and the limitations of product liability directives involved in AVs \\ \hline


Tomlinson \textit{et al.} \cite{tomlinson2018towards} & 2018 & \xmark{} & \checkmark{} & \checkmark{} & \checkmark{} & \xmark{} & \xmark{} &  \checkmark{} & \xmark{} & Presents viable detection methods for the Controller Area Network of cars that are vulnerable to attack \\ \hline


Van \textit{et al.} \cite{van2018survey} & 2018 & \xmark{} & \checkmark{} & \checkmark{} & \checkmark{} & \checkmark{} & \xmark{} &  \xmark{} & \xmark{} & Discusses the Intelligent Transportation System (ITS) and related security issues and presents a classification for security detection mechanisms \\ \hline


Hussain \textit{et al.} \cite{hussain2018autonomous} & 2018 & \checkmark{} & \checkmark{} & \xmark{} & \checkmark{} & \xmark{} & \checkmark{} &  \xmark{} & \xmark{} & Analyzes challenges related to AV development and deployment and highlights applications issues hindering safe and cost-effective AV \\ \hline


Wang \textit{et al.} \cite{wang2018networking} & 2018 & \checkmark{} & \xmark{} & \xmark{} & \checkmark{} & \checkmark{} & \xmark{} &  \xmark{} & \xmark{} & Analyzes autonomous driving from the networking and communication technologies based on intra- and inter-vehicle aspects \\ \hline


Ren \textit{et al.} \cite{ren2019security} & 2019 & \xmark{} & \checkmark{} & \checkmark{} & \checkmark{} & \xmark{} & \xmark{} &  \xmark{} & \xmark{} & Presents an in-depth study on security threats from the angles of perception, navigation, and control and summarizes the corresponding defense mechanisms \\ \hline


Goumidi \textit{et al.} \cite{goumidi2019vehicular} & 2019 & \xmark{} & \checkmark{} & \checkmark{} & \xmark{} & \checkmark{} & \checkmark{} &  \xmark{} & \xmark{} & Discusses and classifies the security challenges related to Vehicular Cloud Computing (VCC) \\ \hline


Al-Jarrah \textit{et al.} \cite{al2019intrusion} & 2019 & \checkmark{} & \xmark{} & \checkmark{} & \xmark{} & \checkmark{} & \xmark{} &  \checkmark{} & \xmark{} & Presents a comprehensive review of the intrusion detection systems with a focus on intra-vehicle aspects  \\ \hline


Sharma \textit{et al.} \cite{sharma2019extended} & 2019 & \xmark{} & \checkmark{} & \xmark{} & \xmark{} & \checkmark{} & \xmark{} &  \xmark{} & \xmark{} & Presents a comprehensive review of security challenges and defense mechanisms for the ubiquitous CAN bus communication protocol \\ \hline


Wang \textit{et al.} \cite{wang2019survey} & 2019 & \checkmark{} & \checkmark{} & \xmark{} & \checkmark{} & \xmark{} & \xmark{} &  \checkmark{} & \xmark{} & Analyzes V2X challenges and requirements and discusses testing methods in the communication process from the architectural perspective \\ \hline


Sommer \textit{et al.} \cite{sommer2019survey} & 2019 & \xmark{} & \xmark{} & \checkmark{} & \checkmark{} & \xmark{} & \checkmark{} &  \xmark{} & \xmark{} & Presents a comprehensive taxonomy of security attacks with details to address them throughout the AV development process \\ \hline


Baldini \textit{et al.} \cite{baldini2020application} & 2020 & \checkmark{} & \checkmark{} & \xmark{} & \xmark{} & \xmark{} & \xmark{} &  \xmark{} & \checkmark{} & Analysis on policy-based framework applications to AVs to regulate, manage, and implement data resources to traffic rules. \\ \hline


Qayyum \textit{et al.} \cite{qayyum2020securing} & 2020 & \checkmark{} & \checkmark{} & \checkmark{} & \checkmark{} & \xmark{} & \xmark{} &  \xmark{} & \xmark{} & Reviews challenges of ML applicability in the communication networks and further formulates the ML pipeline of CAVs along with security threats against ML \\ \hline


Halder \textit{et al.} \cite{halder2020secure} & 2020 & \checkmark{} & \checkmark{} & \xmark{} & \checkmark{} & \xmark{} & \xmark{} &  \xmark{} & \xmark{} & Surveys the security needs for vehicle over-the-air (OTA) software updates \\ \hline


Luo \textit{et al.} \cite{luo2020security} & 2020 & \xmark{} & \checkmark{} & \checkmark{} & \xmark{} & \xmark{} & \xmark{} &  \checkmark{} & \xmark{} & Analyzes security risks of connected vehicles and proposes an evaluation approach to ensure safe system design \\ \hline


Promyslov \textit{et al.} \cite{promyslov2020security} & 2020 & \xmark{} & \checkmark{} & \checkmark{} & \xmark{} & \xmark{} & \xmark{} &  \xmark{} & \checkmark{} & Presents security properties and distributes them into graphs and models for smart AV system and remotely controlled vehicles regarding cybersecurity threats \\ \hline


Pham \textit{et al.} \cite{pham2020survey} & 2020 & \xmark{} & \checkmark{} & \checkmark{} & \xmark{} & \checkmark{} & \xmark{} &  \xmark{} & \xmark{} & Presents an overview of CAV security attacks and a taxonomy of their corresponding countermeasures \\ \hline


Green \cite{green2020self} & 2020 & \checkmark{} & \checkmark{} & \checkmark{} & \xmark{} & \xmark{} & \checkmark{} &  \checkmark{} & \checkmark{} &  Explains the SELF DRIVE Act and NHTSA best safety practices in regard to the regulation of cybersecurity and vehicle hacking in self-driving cars.  \\ \hline


Virag \textit{et al.} \cite{virag2020security} & 2020 & \xmark{} & \checkmark{} & \xmark{} & \checkmark{} & \xmark{} & \xmark{} &  \xmark{} & \checkmark{} & Investigates IT-related security issues for continuously connected vehicles \\ \hline




Wang \textit{et al.} \cite{wang2022security} & 2022 & \xmark{} & \checkmark{} & \checkmark{} & \checkmark{} & \checkmark{} & \xmark{} &  \xmark{} & \xmark{} & Discusses security challenges in CAVs, focusing on sustainability \\ \hline


Taylor \textit{et al.} \cite{taylor2023vehicular} & 2022 & \xmark{} & \checkmark{} & \checkmark{} & \xmark{} & \xmark{} & \xmark{} &  \xmark{} & \checkmark{} & Reviews architectures and security issues in vehicular communication \\ \hline


Hataba \textit{et al.} \cite{hataba2022security} & 2022 & \xmark{} & \checkmark{} & \checkmark{} & \checkmark{} & \xmark{} & \xmark{} &  \xmark{} & \xmark{} & Reviews security and privacy issues in AV from the platooning perspective \\ \hline


Ahmed \textit{et al.}  \cite{ahmed2023vehicular}& 2023 & \checkmark{} & \xmark{} & \xmark{} & \checkmark{} & \xmark{} & \xmark{} &  \xmark{} & \xmark{} & Reviews data offloading, with a focus on inter-vehicle communication network \\ \hline


Deng \textit{et al.} \cite{10184107} & 2023 & \checkmark{} & \xmark{} & \xmark{} & \checkmark{} & \checkmark{} & \xmark{} &  \xmark{} & \xmark{} & This Provides an overview of real-time motion control in vehicles, addressing the impact of network issues like delays, packet dropouts, and congestion from intra-vehicle, inter-vehicle, and integrated perspectives \\ \hline

Wei \textit{et al.} \cite{liu2023systematic} & 2023 & \checkmark{} & \xmark{} &  \xmark{} &  \xmark{} &  \xmark{} & \checkmark{} &  \checkmark{} & \xmark{} & Presents a comprehensive overview of the vehicle control technology, focusing on the evolution from vehicle state estimation and trajectory tracking control in AVs at the microscopic level to collaborative control in CAVs at the macroscopic level.  \\ \hline

Hakak \textit{et al.} \cite{hakak2023autonomous} & 2023 & \checkmark{} & \checkmark{} &  \xmark{} &  \checkmark{} &  \xmark{} & \xmark{} &  \xmark{} & \xmark{} & Provides a comprehensive review on AVs in the 5G and beyond era and discusses on the current advancements in AVs, automation levels, enabling technologies and the requirement of 5G networks. It also presents the impact of 5G and B5G on AVs and the envisaged security concerns in AVs.  \\ \hline
Chellapandi \textit{et al.} \cite{chellapandi2023federated} & 2023 & \checkmark{} & \checkmark{} &  \checkmark{} &  \xmark{} &  \xmark{} & \xmark{} &  \checkmark{} & \xmark{} & Discusses advancements in applying federated learning to connected CAVs, analyzing both centralized and decentralized frameworks and their methodologies, and examining data sources, models, and security techniques crucial for ensuring privacy and confidentiality in federated learning.  \\ \hline
Fang \textit{et al.} \cite{fang2024anomaly}  & 2024 & \checkmark{} & \xmark{} & \checkmark{} & \xmark{} & \xmark{} & \xmark{} &  \checkmark{} & \xmark{} & Considers anomaly detection and interpretation in CAVs. It covers aspects such as defining, classifying, and detecting anomalies, alongside reviewing techniques and advancements, while interpretation includes analyzing anomalies in the context of safety and security risks, ultimately discussing open issues and future directions for anomaly diagnosis in CAVs. \\ \hline
Pipicelli \textit{et al.} \cite{pipicelli2024architecture}  & 2024 & \xmark{} & \xmark{} & \checkmark{} & \checkmark{} & \xmark{} & \xmark{} &  \checkmark{} & \xmark{} & Evaluates the efficiency improvements of CAVs and provides a detailed overview of their architecture, including layout, data processing, and management strategies for energy. It employs statistical distributions in a Monte Carlo simulation to assess the impact of optimal driving behaviors and power consumption of CAV hardware on vehicle energy savings. \\ \hline
Qu \textit{et al.} \cite{qu2024model}  & 2024 & \xmark{} & \xmark{} & \checkmark{} & \xmark{} & \xmark{} & \xmark{} &  \checkmark{} & \xmark{} & Proposes an adaptive cooperative perception scheme for CAV pairs that optimizes between cooperative and stand-alone perception to enhance computing efficiency in mixed-traffic. Utilizing a model-assisted multi-agent reinforcement learning approach, our results demonstrate significant efficiency improvements over traditional methods.\\ \hline
Ibn-Khedher \textit{et al.} \cite{ibn20246g}  & 2024 & \checkmark{} & \checkmark{} & \xmark{} & \xmark{} & \xmark{} & \xmark{} &  \checkmark{} & \checkmark{} & Surveys 6G technologies like mobile edge computing and artificial intelligence that can improve the Internet of Autonomous Vehicles (IoAV) network performance beyond current 5G solutions. We also discuss the convergence of 6G and IoAV for future CAV applications, addressing challenges and proposing directions for 6G-IoAV research.\\ \hline
\textcolor{black}{Mohsen Sorkhpour} \textit{\textcolor{black}{et al.}} \cite{zz7}  & \textcolor{black}{2024} & \xmark{}{} & \checkmark{} & \checkmark{} & \xmark{} & \xmark{} & \xmark{} &  \checkmark{} & \checkmark{} & \textcolor{black}{Control Area Network (CAN) lacks built-in security mechanisms, making it vulnerable to cyber-attacks, and existing Intrusion Detection Systems (IDSs) for CAN require significant manual intervention. Auto-CIDS, an autonomous IDS using Deep Reinforcement Learning (DRL), reduces human intervention by enabling active learning and self-supervised detection of attacks, demonstrating effectiveness on the Car-Hacking dataset}.\\ \hline
\textbf{Our Survey} & \textbf{2024 }& \checkmark{} & \checkmark{} & \checkmark{} & \checkmark{} & \checkmark{} & \checkmark{} &  \checkmark{} & \checkmark{} & \textbf{Discusses the state-of-art security frameworks, challenges, and techniques for securing both inter- and intra-vehicular communication along with related applications and use cases} \\ \hline

\end{longtable}
\normalsize
While this research area is not new, there was a heavy interest in the security of CAVs in the early 2000s. For instance, a group of researchers led by Jean-Pierre Hubaux \textit{et al.} published a series of research papers establishing the groundwork for improving security in CAVs including \cite{4689252}, \cite{5611547}, and \cite{5720206}, and \cite{6336754}. These works include highly detailed designs and frameworks for improving the security of CAVs, designing new mechanisms to improve vehicular communications, and designing protocols for managing communication from multiple pathways for CAVs.

There have been several newer research works that discuss the security of CAVs. Notably, the work \cite{calandriello2010performance} discussed the impact of security on the vehicular communication system, whereas the work \cite{gifei2017integrated} emphasizes the importance of dealing with the security of AVs in an integrative manner and further reviews the importance of establishing acceptable levels for cyber-security risks. Another work \cite{virag2020security} identifies current challenges of securing AVs. The authors conducted interviews with security experts from fields such as consulting, manufacturing, and software engineering and researched topics of failing security in AVs. The work \cite{karnouskos2017privacy} also discusses security for AVs. The authors note that while data plays a crucial role in securing AVs, there are still concerns about privacy and integrity scenarios. Further works such as \cite{baldini2020application} review the Policy-based frameworks (PBFs) for AVs in various applications relevant to security and safety, including network management and access control. Another work \cite{promyslov2020security} discusses the security of CAVs in the context of smart cities, focusing on the integrity of vehicular communications. Furthermore, the work \cite{green2020self} focuses on maintaining security for CAVs while considering the legal aspect, especially regarding users and the design of the CAV. The authors of \cite{tomlinson2018towards} discuss the significance of the automotive CAN in its protection and safety against cyberattacks. The article \cite{parkinson2017cyber} contributes to the cyber threats and automation levels of CAVs and emphasizes knowledge gaps to display vulnerable areas that need improvement in preventing cyberattacks.

Another work \cite{miller2014survey} surveys remote automotive attacks and their surfaces. By expanding on the anatomy of a remote attack, analyzing automotive networks, and presenting ways to defend against remote attacks, the article contributes to preventing cyberattacks in AVs with a focus on ECUs. The article \cite{wang2018networking} expands on the usage of networking and communication devices in AVs, thereby contributing to an understanding of automobile technologies, such as intra- and inter-vehicle communications. The authors investigate and compare the intra- and inter-network connections, communication and networking challenges in CAVs to repair networking and communication issues. Moreover, an overview of CAVs, real-world test results, and the advancements of the technologies that contribute to the expansion of the field is provided in \cite{hussain2018autonomous}. The article focuses on technical and non-technical issues between communications and networking, including software complexity, real-time data analytics, testing, and verification.
Goumidi \textit{et al.} \cite{goumidi2019vehicular} survey cloud computing security in vehicles to underline their importance in VANET improvements. Introducing vehicular ad hoc networks, or VANETs, and combining the technology with vehicular cloud computing (VCC) provides a solid foundation for CAVs. The work \cite{sharma2019extended} highlights vehicle security and presents security measurements to maintain a beneficial driving experience. Security threats and countermeasures for the CAN bus communication protocol are also reviewed in this article. Moreover, the work \cite{al2019intrusion} discusses the different types of intra-vehicle networks that control and monitor the vehicle's state and identifies challenges that the predominant detection systems cannot address.

Additional surveys such as \cite{pham2020survey} present security attacks and defense mechanisms for CAVs. In this article, security attacks are quantified and compartmentalized along with their corresponding countermeasures on CAVs. Additionally, the article \cite{qayyum2020securing} provides a comprehensive overview of techniques and challenges related to the application of ML in CAV networks. However, different from these two articles, our work discusses the state-of-art security protocols, architectures, and frameworks in addition to security challenges and implications, focusing on intra- and inter-communication.

Interestingly enough, mobile car apps for CAVs have become increasingly popular in recent years due to their benefits:  convenience, control, and visibility to the users. However, mobile car apps for CAVs add additional risks: exposing vulnerabilities in the mobile app, the phone's operating system, and the app's usability. Additionally, sometimes vendors are eager to release a new mobile app for CAVs while overlooking security features. As such, these car apps for CAVs are often released and updated without necessary security requirements. Furthermore, mobile car apps for CAVs are challenging due to the complexity of the CAV ecosystem with mobile environments. Its complex data architecture makes it difficult to secure communications between multiple sources: the mobile phone, the mobile application, the application's servers, the connected vehicle itself, and the servers communicating with the car. Thus, To protect the multiple sources, it is not enough to secure each of them separately – the data between them must be correlated and analyzed to produce a coherent view of the data flow. There must be a network-based solution to secure the entire data center and the network \cite{zz1}.

Several studies have covered mobile car apps. For example, Eriksson \textit{et al.} \cite{eriksson2019road} explored on the Android automotive system architecture with a special focus on the permission model. The authors also addressed several vulnerabilities, implemented several attacks on the android automotive system, and measured the severity of these attacks using the Common Vulnerability Scoring System (CVSS3). It was found that the biggest vulnerabilities come from Android's permission model. In \cite{mandal2018vulnerability}, the authors also focused on Android and examined car apps available on Google Play and found that nearly 80\% of the apps are potentially vulnerable. Similar findings were also discovered in works \cite{mandal2019static} and \cite{lee2019enhanced}.

The works of \cite{mandal2018static} and \cite{chatzoglou2021multi} also discuss how using automotive smartphone apps provided by car manufacturers can offer numerous advantages to the vehicle owner, such as improved safety, fuel efficiency, monitoring of vehicle data, and timely software updates. However, the continuous tracking of the vehicle data by such apps may also pose a risk to the CAV owner. While the work \cite{mandal2018static} explores the OpenXC platform (an open-source API that allows Android apps to interact with the car's hardware) and shows how it is used to create interjection attacks, the work \cite{chatzoglou2021multi} assesses the security of all official single-vehicle management apps offered by major car manufacturers who operate in Europe. It is found that many apps remain susceptible to critical vulnerabilities due to misconfigurations, secure coding negligence, and latent or overt security flaws that relate to the use of third-party software. \textcolor{black}{Further, the book by Singh \cite{madhusudan2021information} presents a cybersecurity overview and perspectives along with challenges that impact vehicular technology. It also discusses security structures and potential solutions to specific security issues. } \textcolor{black}{Additionally, the work \cite{zz7} introduces Auto-CIDS, an autonomous IDS for CAN networks that leverages deep reinforcement learning to enhance intrusion detection, significantly reducing the need for manual intervention.}

As the aforementioned works reveal, security mechanisms must be considered fully for the CAVs. Furthermore, both technical and non-technical aspects---such as security implications, impact assessments, security evaluation metrics, and regulatory issues---must also be considered. While the works discussed in this section may arguably seem similar to some, they are notable in terms of the areas they address and potential future trends for CAVs.

\subsection{Contributions and Paper Organization}

\textcolor{black}{The complex digital ecosystems in CAVs introduce unique security vulnerabilities across sensors, controllers, and communication interfaces. These security concerns, \textcolor{black}{ranging from potential vehicle hijacking to traffic data manipulation,} directly impact public safety, trust, and the broader economic potential of CAVs. Thus, this review \textcolor{black}{is essential in addressing and understanding these challenges} \cite{zz2,zz3}. It aims to foster public trust, guide regulatory frameworks, highlight research gaps in communication security, and help balance rapid technological advancements \textcolor{black}{with safety assurances}.} 

\textcolor{black}{While numerous related surveys exist in the literature, none bridges the gap between intra- and inter-vehicular communication attacks. This gap arises because: 1) Covering both areas requires a comprehensive understanding of multiple disciplines, such as cybersecurity, automotive engineering, and wireless communications; 2) Addressing both intra- and inter-vehicular communication in a single review is intricate due to the range of components and communication protocols involved. \textcolor{black}{Researchers often find it more manageable to focus on one aspect rather than both;} and 3) The integration of CAVs into mainstream transportation is still relatively new. As technologies evolve and new research emerges, up-to-date comprehensive reviews will remain essential for researchers and practitioners. This paper is the first to systematically synthesize both intra- and inter-vehicular attacks and defenses, \textcolor{black}{making a significant contribution to the existing body of literature summarized in Table \ref{tab:survey}.}} 

This paper makes significant contributions to the field of CAV communications, \textcolor{black}{which are detailed as follows:}
\begin{itemize}
  \item \textcolor{black}{We present a thorough outline of cutting-edge protocols recently introduced in scholarly literature, emphasizing their advancements and applications in CAVs.}
  \item \textcolor{black}{The paper offers a comprehensive examination of prevalent CAV architectures, providing in-depth insights into each architecture's structure, functionality, and impact on CAV operations.}
  \item We propose a set of practical security protocols designed for easy integration into existing CAV systems. These protocols enhance data integrity and confidentiality without imposing significant overhead on system resources, striking an optimal balance between security and performance.
  \item Unlike prior studies \textcolor{black}{that} focus on isolated security threats, our work offers an exhaustive analysis of both intra- and inter-vehicular attack vectors. We categorize these attacks and demonstrate through simulations how they impact CAV operations, providing a holistic view that aids in better security planning and system design.
  \item A variety of use cases are explored to demonstrate the practical applications of the discussed protocols. Additionally, the paper incorporates insights from academic researchers, industry professionals, and international regulatory bodies, offering a multidimensional understanding of CAV architectures.
  \item We identify and propose potential areas for future research, focusing on unexplored aspects of CAV communication security and technology development. This includes suggestions for advancing current protocols and architectures, exploring new security challenges in evolving CAV ecosystems, and developing innovative solutions to enhance the safety and efficiency of autonomous vehicles.
  \item Our study leverages interdisciplinary approaches, combining insights from cybersecurity, automotive engineering, and wireless communications to forge comprehensive solutions that are robust against a wide spectrum of potential vulnerabilities.
\end{itemize}



\textcolor{black}{The paper is organized into eight comprehensive sections. 
Specifically, the structure is as follows: Section 2 provides background on communication security in CAVs, and Section 3 reviews communication security architectures and challenges. Next, Section 4 presents an up-to-date taxonomy of communication security attacks on CAVs. While Section 5 discusses state-of-the-art communication security solutions and evaluation tools, Section 6 explores communication security protocols for CAVs and best practices. Section 7 then provides in-depth insights on open issues and future research directions. Last, Section 8 concludes the article.}

\section{Background on Communication Security}
\textcolor{black}{This section presents key use cases and applications involving communication security in autonomous driving, followed by a discussion on standards related to CAV security.}

\subsection{Autonomous Driving Applications that Demand Security}
\textcolor{black}{The need for efficient security mechanisms in CAVs has driven researchers to develop specific use cases that highlight their approaches to securing various applications of CAVs} \cite{oehl2020affective, zz8}. Key applications of autonomous driving will be discussed next.

\subsubsection{Valet Parking}
\textcolor{black}{Secure communication plays a crucial role in the valet parking use case for CAVs \cite{chehri2019autonomous}. One key aspect is establishing a secure channel between the CAV and the parking infrastructure, ensuring that sensitive information—such as the vehicle's identity, location, and authorization credentials—is protected from eavesdropping and tampering by malicious actors \cite{salek2022review}. Secure communication is particularly important during the initial authentication and authorization process, where the CAV securely identifies itself with the parking system and obtains the necessary permissions to park. Additionally, secure communication must be maintained throughout the parking duration to enable continuous monitoring and control over the vehicle's status and prevent unauthorized access. By employing robust encryption algorithms, authentication mechanisms, and secure protocols, CAVs can safeguard communication with the parking infrastructure, minimizing the risk of unauthorized access and ensuring a secure valet parking experience \cite{9447840}.}

Driverless parking is one of the benefits of CAVs that has gained notable attention due to the increased focus on designing automated and secure parking models \cite{pokhrel2020privacy, iroshnikov2020autonomous}. \textcolor{black}{However, location and identity privacy issues persist because parking data is sometimes shared improperly. To address this, a privacy-centric framework for secure parking reservations has been proposed} \cite{scholliers2020automated}. Additionally, an Autonomous Valet Parking System (AVPS) uses connected communication technologies to locate available parking spots in garages and park the vehicle autonomously \cite{khalid2020toward}. \textcolor{black}{A critical factor emphasized in \cite{khalid2020toward} is ensuring that the reserved parking space remains available until the CAV is parked or the timer expires, which can increase the risk of a Multiple Reservation Attack (MRA).} Under this attack, the user attempts to reserve all other possible parking spaces, assuming other users have already reserved them. Furthermore, the registration process required to secure a parking spot may inadvertently expose user identity and personal data, such as unique identifiers, email addresses, and home addresses.

\subsubsection{Lane Changing}
Lane changing is another prominent use case of autonomous driving applications \cite{fu2020autonomous}. After all, inappropriate lane following and changing behaviors of CAVs can lead to accidents. Using deep-reinforcement learning as a widely used approach to ensure this does not occur \cite{liu2020driving}. However, there are still issues with validating reinforcement learning systems for autonomous systems. This can lead to incoherent interpretations of the performance of the reinforcement learning algorithms and how well they generalize \cite{ferdowsi2018robust, fu2020autonomous}. Other issues with deep-reinforcement learning are sample efficiency, difficulty in reproducing reinforcement-learning results, and high sensitivity to hyper-parameter choices \cite{xia2016control, sallab2017deep}. As such, a blockchain-based learning (BCL) framework for autonomous lane-change systems was proposed in \cite{fu2020autonomous}. In the BCL architecture, four main issues are addressed: learning rate, data security, users' privacy, and communication burden while achieving a success rate of 96.5\% for lane changes, with the likelihood of accidents or failures occurring ranging between 1\% to 2\%.

\textcolor{black}{When a CAV is performing a lane change maneuver, it needs to communicate with other vehicles and the surrounding infrastructure to ensure a safe and efficient transition \cite{ma2023collision}. Secure communication comes into play at various stages of this process. Initially, the CAV needs to securely broadcast its intent to change lanes, informing nearby vehicles of its planned action. This communication ensures that other vehicles can adjust their behavior accordingly to maintain a safe distance and avoid collisions. Further, secure communication is critical for the CAV to receive feedback from surrounding vehicles, confirming their acknowledgment and cooperation \cite{ghorai2022state}. Throughout the lane change maneuver, continuous secure communication is necessary to exchange real-time information, such as the CAV's position, speed, and trajectory, with other vehicles and the traffic management infrastructure. }

\subsubsection{Image and Audio-based AI Service for Security Applications}
Image and audio-based approaches can incorporate various AI (Artificial Intelligence) methods to increase security and trust inside an autonomous shuttle \cite{tsiktsiris2022novel}. This approach runs in real-time and can be used to detect petty crime scenarios (e.g., screaming, bag snatching, people fighting, vandalism). This system can further enable notifications to authorized personnel so they can dictate proper actions. Thirteen different scenarios were simulated using different camera perspectives for each one, and 29 video sequences were also captured with 46,127 frames, achieving up to 96\% accuracy for target classes classification \cite{tsiktsiris2022novel}.
\textcolor{black}{When image and audio-based AI services are deployed for security applications in CAVs, such as detecting and responding to suspicious activities using image and audio analysis, secure communication becomes crucial. Firstly, the CAV must securely transmit the captured image and audio data to the central security system for analysis \cite{mudhivarthi2023aspects}. This involves employing encryption techniques and secure protocols to protect the confidentiality and integrity of the data during transmission, ensuring that sensitive information is not compromised. Additionally, secure communication is critical when receiving instructions or updates from the central security system. This ensures that the CAV can authenticate and verify the authenticity of the commands it receives, preventing unauthorized access or tampering.}

\subsubsection{Web-Based Car Control and Monitoring for Safe Driving}
The security of vehicular communications, especially the V2I (vehicle-to-infrastructure) and V2V (vehicle-to-vehicle) communications is considered to enhance autonomous driving security and experience. A web-based car monitoring was proposed to enable vehicles in vehicular networks to exchange their mobility information, achieving quick road awareness using W3C Vehicle Information Service Specification (VISS) and Vehicle Signal Specification (VSS) \cite{9687175}. This use case takes advantage of the vehicle network infrastructure to control and monitor road mobility to enhance the safety of autonomous driving. The implemented use case typically incurs a delay of 0.055 seconds for V2I communication and an average delay of 0.0015 seconds for V2V, enabling distant CAVs to proactively react. 

\subsubsection{Blockchain-inspired Event Recording System}
What is unique about \cite{8606016} is that this particular use case covers the liability aspect of security. When CAVs are involved in accidents among themselves or with human subjects, liability must be based on accident forensics. Thus, a blockchain-based event recording system was proposed for CAVs \cite{8606016}. A new PoE (Proof of Event) mechanism was also designed to achieve indisputable accident forensics by ensuring that event information is trustworthy and verifiable. Along with the event recorder system, the consensus mechanism efficiently verifies and confirms new blocks of event data without any central authority. Additionally, accidents are recorded as time-stamped transactions that are supposed to be saved into new blocks in real-time. A cellular network-based infrastructure is adopted to define a vehicular network for each accident.

\textcolor{black}{
Secure communication comes into play in this use case when the CAV transmits the event data to the blockchain network for recording \cite{yazdinejad2024secure}. This may involve encrypting the data to protect its confidentiality and integrity during transmission, preventing unauthorized access or tampering. Further, secure communication is necessary when verifying and validating the event data with other network participants. This ensures that the recorded events are authenticated and cannot be tampered with, maintaining the integrity and immutability of the blockchain ledger.}

\subsection{Standards Relating to CAV Communication Security}
For this subsection, we focus on relevant standards related to the communication security of CAVs. Standards developed for CAVs often cover multiple aspects, such as functional safety, operation correctness, etc. These standards are not nearly talked about enough, likely due to the complexities of the CAV itself and the challenges of developing threat models to ensure maximum testing of the CAV.

The unfortunate reality is that threat modeling is challenging, and it can be used as a tool to aid software and algorithm design, as well as to help design teams think more clearly about what types of security issues they may face. However, without real-world experience of how systems are attacked at an engineering level, it can be far too abstract. Therefore, It is essential that those involved in forming standards for CAVs and the teams involved truly understand the types of threat environments and help formulate guidelines to cover as much basis as possible. This is because a threat or security risk scored too highly can increase the development cost. Conversely, a score that is too low will starve funds and engineering resources from an area that sorely needs attention. Numerous standards have been designed to address these concerns, including the International Organization for Standardization (ISO) 26262 SAE J3061, among others, which are detailed next.
\textcolor{black}{Communication security is a critical aspect of CAVs, and it is typically addressed through specific standards and frameworks that complement baseline standards such as ISO 26262. Prominent, relevant standards and frameworks that specifically address communication security in CAVs are summarized next.}

\subsubsection{SO/SAE 21434}
\textcolor{black}{This standard, also known as Automotive Cybersecurity Engineering, provides guidelines for automotive cybersecurity throughout the vehicle's lifecycle. It addresses various aspects of cybersecurity, including secure communication protocols, secure software development, risk assessment, and vulnerability management \cite{taylor2023vehicular}.}

\subsubsection{SAE J3061} 
\textcolor{black}{Developed by the Society of Automotive Engineers (SAE), this document provides a cybersecurity guidebook for the automotive industry. It covers the entire vehicle ecosystem, including communication security, threat analysis, risk assessment, incident response, and security lifecycle management.}
\textcolor{black}{This particular standard is based on the ISO 26262 standard and covers the growing threat landscape of CAVs. SAE J3061 establishes the terminology of threat, vulnerability, and risk, providing an overview and distinction between system safety and system cybersecurity. It also presents a few general guiding cybersecurity principles that apply to any organization within a company. Another notable component is that this standard provides information regarding some common existing tools and approaches for the design, verification, and validation of CAVs. A foundation is presented for further development approaches for CAV cybersecurity. The most important contribution this standard offers is this distinction: that a cybersecurity-critical system may not be safety-critical \cite{9447840}.}

\subsubsection{AUTOSAR Security} 
\textcolor{black}{The AUTOSAR (Automotive Open System Architecture) consortium defines a set of standards and specifications for automotive software architectures. Within AUTOSAR, security-related standards and mechanisms are defined, including secure communication protocols, secure bootstrapping, secure onboard communication, and secure software updates \cite{benyahya2022automated}.}

\subsubsection{NIST Cybersecurity Framework} 
\textcolor{black}{Although not specific to the automotive industry, the National Institute of Standards and Technology (NIST) Cybersecurity Framework offers a comprehensive set of guidelines and best practices for managing and improving cybersecurity risk. This framework can be applied to CAVs to address communication security aspects \cite{sheehan2019connected}.}

\subsubsection{ISO 21434}
The ISO 21434 standard guides cybersecurity risk management by specifying engineering requirements relating to cybersecurity risk in terms of concept, product development, operation, maintenance, and decommissioning. The standard also covers various components and interfaces. Additionally, the ISO 21434 standard includes a framework that includes a common language for communicating and managing cyber security risk \cite{9447840}. \textcolor{black}{ISO 21434 is intended to provide guidelines and requirements for incorporating cybersecurity into the engineering processes of road vehicles. It is expected to address various aspects of automotive cybersecurity, including communication security.}

\subsubsection{ISO 20078-3:2021}
The focus of this standard is on user authentication. Specifically, it describes how to authenticate users and other parties on a web service interface. A definition of how a resource owner can delegate access to a trusted party is also provided. The standard also defines the necessary roles and crucial separation of duties to fulfill tasks relating to security, data privacy, and data protection. A reference implementation using an OAuth 2.0 compatible framework and OpenID Connect 1.0 compatible framework is also provided \cite{1400-1700_iso_nodate}.

\subsubsection{ETSI TR 103 460}
This standard provides an overview of relevant misbehavior detection mechanisms suitable for CAVs. The standard also provides information regarding performance and applicability on different misbehavior mechanisms and describes minimum requirements for security architecture \cite{ansari2021v2x}. \textcolor{black}{Further, ETSI TR 103 460 provides guidance on the security architecture and governance framework for Cooperative Intelligent Transport Systems (C-ITS) systems. It addresses the specific security challenges and requirements associated with CAV communication and aims to ensure secure and trusted communications between vehicles, infrastructure, and other entities in the C-ITS ecosystem.}
\textcolor{black}{It is worth noting that baseline standards such as ISO 26262 and other related standards and frameworks are not mutually exclusive. For instance, while ISO 26262 primarily focuses on functional safety, cybersecurity, and communication security are essential components of overall safety in CAVs. Therefore, organizations developing CAV systems typically consider ISO 26262 and other relevant standards to address both functional safety and communication security aspects effectively.}

\section{Communication Security Architectures and Challenges} \label{securityArchitecture} 
This section presents the prominent CAV security architectures along with key communication security challenges.

\subsection{Security Architectures for CAVs}

\subsubsection{RACE}
The authors of \cite{yuan2020race} propose a security framework called RACE (reinforced cooperative autonomous vehicle collision avoidance) geared toward collision avoidance. CAVs typically gain proficiency with their general surroundings and ideal driving procedures via onboard sensors like Lidar, radar, and acceleration sensors. In any case, due to the restrictive communication range of the vehicles, the gathered data from one vehicle is inadequate to satisfy enormous large-scale road safety requirements and improve collision performance efficiently. As such, the authors designed the RACE architecture to overcome these shortcomings. The authors built their framework using a simulator and evaluated their framework on the simulator as well. Upon examining the evaluation results, the authors concluded their framework can learn from the environment and its neighboring agents. RACE also managed to outperform other frameworks and reduce collisions by 65\%.

\textcolor{black}{One of the key innovations of RACE is its utilization of cooperative perception, where data from multiple vehicles is combined to enhance situational awareness. This collaborative approach allows vehicles to have a broader view of their surroundings beyond the limitations of their own sensors, significantly improving decision-making accuracy. Additionally, the framework incorporates reinforcement learning algorithms, which adapt the behavior of vehicles based on past experiences, allowing the system to progressively optimize collision avoidance strategies. However, further analysis is required to assess how well the system performs in highly congested environments and under scenarios involving unexpected road events, such as sudden pedestrian crossings. This deeper exploration can provide insights into the scalability and real-world applicability of RACE.}

\subsubsection{Blockchain Frameworks for Securing CAVs}
Recently, many researchers are incorporating blockchain frameworks for CAVs \cite{singh2017blockchain, dorri2017blockchain, narbayeva2020blockchain}. Blockchain is a popular and viable choice for securing CAVs due to its high security and immutability. The work by Rathee \textit{et al.} \cite{rathee2019blockchain} proposes a blockchain architecture for securing CAVs that emphasizes secrecy and protection and is presented as a use case for service delivery. For the proposed framework, CAVs are registered into a network before services can be accessed. The required information on the CAVs is entered into a database first, then stored on the blockchain-centric framework permanently so that every event can be accounted for in the CAVs. The framework also involves IoT devices (Internet of Things), which are also connected to CAVs. The quantity of CAVs connecting to those IoT devices depends on communication and transmission ranges. In addition, To account for the likelihood of the CAVs or devices being hacked, every device that provides the information about the CAVs needs to register itself on the network before providing services to the connected vehicles. This security framework leverages blockchain capabilities to enforce both privacy and secrecy for vehicle drivers by recording all activities and tracing them inside the blockchain ledger \cite{s3}. The extracted IoT data is further recorded and inspected by various authorities to ensure the driver's safety and the IoT device's security.

\textcolor{black}{However, while blockchain offers high levels of immutability and security, the inherent challenge of scalability and latency remains unresolved. Specifically, the time required to process and validate transactions within a blockchain network may hinder real-time vehicular communication, which is critical in safety-sensitive CAV environments. For instance, scenarios such as rapid lane changes or emergency braking require near-instantaneous data sharing across the network. Implementing off-chain solutions or hybrid blockchain architectures could mitigate these latency issues. Furthermore, the computational overhead for resource-constrained IoT devices involved in blockchain verification processes must be optimized to ensure seamless integration into CAV ecosystems. Future work could focus on lightweight consensus algorithms to reduce the burden on these devices while maintaining security guarantees.}

Blockchain technology can also be leveraged to handle information access to restricted entities within the CAV ecosystems and authentication mechanisms for fog computing-enabled Internet of Vehicles \cite{eddine2021easbf}. For instance, a challenge/response-based exchange of information between roadside units (RSUs) and connected vehicles can be leveraged to monitor the internal state of vehicles and detect intra-vehicle security compromise events. Furthermore, through blockchain technology, valid and legitimate communication traffic accuracy can be enhanced within the intra-vehicle network by assuring only vehicles with a verifiable record in the public ledger can exchange messages. A remarkable solution is presented in \cite{oham2021b}, where the authors propose a Blockchain-based Framework for sEcuring smaRt vehicLes (B-FERL), a permissioned blockchain-based framework, to regulate information access in the CAV ecosystem. Here, quantitative-based emulation scenarios demonstrate the efficiency of B-FERL in ensuring reliable trust management and response time to vehicle network attack identification.

\subsubsection{Privacy-Preserving Raw Data Sharing Framework}
Xiong \textit{et al.} \cite{xiong2020edge} focus on establishing a security architecture for CAVs, but their framework is more focused on data sharing, data leakage, and encryption. Because CAVs will often share data, protection must be provided to prevent possible data leakage \cite{wei2016trinc, hussain2015covert}. However, encrypting data often leads to tremendous overhead being generated, and even with current research, this is still a persistent issue. To alleviate these issues, the authors implement a privacy-preserving raw data-sharing framework using Edge Assist. The authors note that because their architecture incorporates edge computing, their proposed architecture is lightweight and privacy-preserving.
\textcolor{black}{A significant technical challenge in this approach is balancing data privacy with real-time processing efficiency. By offloading data processing to edge servers, the framework reduces the computational burden on CAVs; however, edge nodes themselves must remain highly secure and capable of handling large volumes of encrypted data without introducing latency. Another aspect worth deeper investigation is the framework’s resilience to adversarial attacks targeting the edge servers. These servers could become central points of failure if compromised, leading to widespread data breaches across connected vehicles. A possible solution is the integration of distributed edge computing models, where no single node has complete control over the data, thereby reducing the attack surface. Furthermore, analyzing the trade-offs between encryption strength and computational overhead in various traffic conditions will be critical for determining the framework’s scalability in real-world applications.}

In this proposed framework, the architecture consists of CAVs, edge-based servers, and an overall main trusted server. A CAV has to understand its surrounding environment by using different sensors and generating real-time sensor data. Each captured image is encrypted into two encrypted images, called ciphertexts. The two ciphertexts are transferred to two edge servers, both implementing a deep learning (DL) framework for image processing. Another component involved in the author's proposed framework is a Privacy-Preserving Convolutional Neural Network (P-CNN). The P-CNN handles object classification. The authors used the VGG-16 CNN model and modified it to design the P-CNN. Afterward, the authors tested their proposed framework on the KITTI dataset. Upon examining the results, the proposed framework was able to reduce computational costs and overhead without leaking private data.

\subsubsection{AC4AV}
In another work \cite{zhang2020ac4av}, Zhang \textit{et al.} propose an architecture called AC4AV, which revolves around securing the sensing data of AVs. After all, sensing data allows AVs to perceive their environments, and tampering with this data leads to inaccurate driving decisions while also threatening passenger safety \cite{xu2017raptor}. Due to how complex the AVs' sensing data is, establishing a secure framework is challenging. There is currently no suitable framework that can solve this issue, and current architectures only focus on one element. This is why the authors proposed the AC4AV architecture: to provide a flexible and dynamic way of implementing security for AVs. As such, AC4AV focuses on access control, which is supposed to shield information from fraudulent access.
\textcolor{black}{A key strength of AC4AV is its adaptability, particularly in how it can adjust access control policies based on the dynamic nature of AV environments. The framework's context monitoring component continuously updates the system's understanding of its surroundings, which allows for real-time adjustments to access permissions. However, there is a need for further analysis on how AC4AV handles conflicting access control decisions, especially in scenarios where multiple entities (e.g., passengers, external devices) request access simultaneously. Additionally, the framework’s reliance on predefined access policies may limit its ability to handle novel security threats. Enhancing the system with machine learning techniques could enable it to detect and adapt to new attack vectors that have not been explicitly accounted for in the access control engine. Performance evaluations of AC4AV in diverse real-world conditions, such as urban and rural environments, would provide valuable insights into its robustness and scalability.}

Regarding the AC4AV architecture, the Access Control Engine has four main components: access enforcement, context monitoring, data abstraction, and engine API. The context monitor obtains the CAV system's status information. Next, the data abstraction component's goal is to make the data from CAVs easier to comprehend and distinguish. The Engine API component is meant to support a customized policy regarding security. The Engine API also has several implemented access control modules so that various programs can assign different access control models to protect their information. Finally, the access control engine is in charge of ensuring that permissions to access various actions for CAVs are accurately verified. In terms of evaluation, the authors evaluated AC4AV based on performance and concluded that AC4AV reduces latency and has low overhead. There is also little impact on real-time data access.

\subsubsection{ML-Based Detection for Cybersecurity Attacks on CAV}
In \cite{he2020machine}, the authors highlight that there is no standard framework for securing CAVs, especially regarding some attacks that CAVs are prone to. The work explicitly centers around the network safety aspect of self-governing vehicles. The authors propose a Unified Modeling Language (UML)-based cybersecurity framework and utilize that framework to identify and distinguish other possible vulnerabilities of CAVs. In the framework, the authors also developed a new data set for CAVs named CAV-KDD, which is based on the KDD99 dataset. The CAV-KDD data set has 14 communication-based sub-attacks. The authors also used the Decision Tree and Naive Bayes classifiers for the new dataset and compared their performances based on accuracy, precision, and runtime. For the authors' experiment, the authors split their data between normal data and malevolent data for training and testing. To avoid possible over-fitting, the authors used 10-fold validation. Upon examining the experiment results, the authors note that the decision tree model achieved the best results and is suitable for detecting various types of cyberattacks on CAVs. However, both ML models performed poorly when tested on new types of unseen CAV cyber-attacks.
\textcolor{black}{One of the major challenges identified in this work is the poor performance of ML models when faced with unseen attack types. This is particularly problematic in the highly dynamic environment of CAVs, where new attack strategies emerge frequently. Future work should focus on developing adversarial training methods to improve the generalizability of ML models across a wider range of potential threats. Moreover, the authors could explore federated learning as an alternative approach, where models are trained collaboratively across multiple CAVs without sharing sensitive data. This would allow the system to learn from a broader dataset while preserving privacy and increasing robustness against diverse attack patterns. The trade-offs between model accuracy and computational efficiency must also be analyzed, as resource-constrained CAVs may struggle to run complex models in real-time.}

Additionally, a considerable challenge here is the design of an efficient autonomous controller to carry out real-time control-based autonomous driving decisions (e.g., acceleration level when merging or changing lanes on the highway). Existing frameworks and prototypes mainly rely on conventional feedback controllers, which are traditional ML-based controllers trained through the local data of CAVs. Therefore, such solutions may degrade driving safety by providing inaccurate decisions over the various traffic dynamics and road conditions \cite{zeng2021federated}. A notable work carried out by Zeng \textit{et al.} \cite{zeng2021federated}, where they propose a novel dynamic federated framework (DPF) leveraging a proximal learning algorithm. The authors' proposed solution accounts for the CAV wireless fading channels, unbalanced distributed data across multiple vehicles, and CAV mobility conditions. The proposed solution uses at-scale-based connectivity to develop the CAV autonomous feedback controller, where the controller learning and training models are collaboratively carried out over multiple CAVs to provide optimal decision convergence.

Software Defined Networking (SDN) technology supports efficient communication within the data plane and management plane (through the OpenFlow protocol) \cite{s1,s2}. It further enables executing complex traffic flow transmission rules using inexpensive and simplified forwarding devices such as virtual software switches (e.g., Open vSwitch). Security, time-sensitive forwarding, and easily updatable network devices can benefit from integrating SDN into autonomous driving systems. To support this, Hackel \textit{et al.} \cite{hackel2019software} discussed the potential integration of SDN technology and Time Sensitive Networks (TSN) into intra-vehicular networks. The authors proposed a Time-Sensitive Software-Defined Networking (TSSDN) framework that deploys time-sensitive traffic over SDN-enabled switches by registering data flows with OpenFlow. The proposed framework was evaluated with respect to real-time capabilities on a standard event-based network simulator, namely, OMNeT++. Last, the authors concluded that although SSSDN frameworks add control computational overhead and latency, no latency penalty applies to the TSN traffic.

\subsubsection{5G-V2X}
\textcolor{black}{The 5G-V2X architecture has been developed by the 5G Automotive Association (5GAA), which is a corporate coalition to design and develop standardized architectures and protocols for vehicular technology using 5G-based communications \cite{yazdinejad2019blockchain}. 5GAA has developed the 5G-V2X architecture, which focuses on leveraging 5G technology for CAVs \cite{abdel2020current}. The architecture encompasses the following key components and functionalities:
\begin{itemize}
    \item \textbf{Radio Access Network (RAN)}: The RAN component includes the 5G base stations that provide wireless connectivity for CAVs. It enables direct communication between vehicles, infrastructure, and other devices through the 5G air interface.
    \item \textbf{Cellular Core Network}: The cellular core network manages the communication between CAVs and other entities within the 5G network. It includes various elements, such as the user plane function (UPF) and the control plane function (CPF), that handle data forwarding and control signaling, respectively.
    \item \textbf{V2X Communication}: The 5G-V2X architecture supports Vehicle-to-Everything (V2X) communication, which involves data exchange between vehicles, infrastructure, pedestrians, and the cloud. This communication enables real-time information sharing related to traffic conditions, road hazards, and cooperative maneuvers.
    \item \textbf{Edge Computing}: Edge computing is integrated into the 5G-V2X architecture to enable low-latency processing and decision-making closer to the network edge. By deploying computing resources near the RAN, edge computing enhances the responsiveness of CAV applications that require real-time data processing \cite{yazdinejad2023accurate}.
    \item \textbf{Cloud Connectivity}: Cloud platforms play a vital role in the 5G-V2X architecture by providing storage, processing capabilities, and data analytics for CAVs. The cloud allows for functions such as remote monitoring, firmware updates, and advanced data analysis, contributing to enhanced decision-making and personalized services.
    \item \textbf{Security and Privacy}: The 5G-V2X architecture prioritizes robust security mechanisms to protect the privacy of users and ensure the integrity and confidentiality of the data exchanged. Encryption, authentication, access control, and intrusion detection systems are typically incorporated to secure CAV systems.
\end{itemize}}

\textcolor{black}{While the 5G-V2X architecture offers significant advancements in terms of communication speed and reliability, one of the critical challenges lies in maintaining consistent low-latency communication in highly congested urban environments. The density of vehicles and infrastructure in such settings may result in interference, which can degrade communication quality and delay real-time decision-making processes. Additionally, the integration of edge computing is promising for reducing latency, but it introduces the need for robust and scalable infrastructure at the network edge to handle the massive volumes of data generated by CAVs. Further research is required to explore the optimization of edge node distribution and resource management, particularly in rural or less-developed areas where infrastructure may be sparse.
Another challenge is the security aspect, where the use of 5G networks increases the attack surface due to the higher number of connected devices and distributed components. While encryption and authentication mechanisms are integral to the architecture, the complexity of managing security across multiple layers, from the edge to the cloud, requires advanced intrusion detection systems that can adapt to evolving threats. Moreover, the security of V2X communication must account for the varying trust levels among different communication entities, such as pedestrians, vehicles, and infrastructure, which could introduce vulnerabilities if not properly managed. Implementing zero-trust security models within 5G-V2X environments could enhance overall security by ensuring that every entity is continuously authenticated and verified before communication.}

\subsection{Communication Security Challenges}

Even with current research and progress regarding CAVs, there are still many challenges that persist, especially concerning security \cite{al2021addressing}. The challenges include privacy, real-time response, data validation, malware detection, and liability, as shown in Table \ref{tab:AV-Challenge-Security-Components}.

\begin{table*}[!t]
\centering
\tiny
\caption{An overall summary of AV components, their tasks, attacks/risks they are vulnerable to, and the potentially compromised security goals.}
\label{tab:AV-Challenge-Security-Components}
\resizebox{\textwidth}{!}{%
\begin{tabularx}{\textwidth}{|X|X|X|X|X|}
\hline
\rowcolor[HTML]{DAE8FC} 
\textcolor{black}{\textbf{CAV Component}} & \textcolor{black}{\textbf{Area(s) of Focus}} & \textcolor{black}{\textbf{Attack(s)}} & \textcolor{black}{\textbf{Compromised Security Goals}} & \textcolor{black}{\textbf{CAV Challenges}} \\ \hline
\textcolor{black}{GPS} & \textcolor{black}{Navigation} & \textcolor{black}{Jamming, Spoofing} & \textcolor{black}{Data integrity, Data confidentiality} & \textcolor{black}{Real-time Response} \\ \hline
\textcolor{black}{LIDAR} & \textcolor{black}{Sensing} & \textcolor{black}{Relay, Spoofing} & \textcolor{black}{Data integrity, Data confidentiality, Authentication} & \textcolor{black}{Privacy, Real-time Response} \\ \hline
\textcolor{black}{MMW Radar} & \textcolor{black}{Warning System} & \textcolor{black}{Jamming, Spoofing} & \textcolor{black}{Data integrity, Data confidentiality} & \textcolor{black}{Liability} \\ \hline
\textcolor{black}{Ultrasonic Sensors} & \textcolor{black}{Park Assist} & \textcolor{black}{Spoofing, Jamming} & \textcolor{black}{Data integrity, Data confidentiality} & \textcolor{black}{Malware Detection} \\ \hline
\textcolor{black}{Camera(s)} & \textcolor{black}{Visual representation of Environment} & \textcolor{black}{Blinding} & \textcolor{black}{Integrity, Availability} & \textcolor{black}{Real-time Response} \\ \hline
\textcolor{black}{OBD Port} & \textcolor{black}{Real-time data} & \textcolor{black}{Malicious OBD devices, Remote exploitation} & \textcolor{black}{Data integrity, Availability} & \textcolor{black}{Malware Detection} \\ \hline
\textcolor{black}{ECU} & \textcolor{black}{Controlling other sub-systems} & \textcolor{black}{ECU access through CAN, ECU access via other connection mechanisms} & \textcolor{black}{Data integrity, Availability} & \textcolor{black}{Liability, Real-time Response, Malware Detection} \\ \hline
\textcolor{black}{CAN} & \textcolor{black}{Communication} & \textcolor{black}{Malicious attacks, Passive Attacks, Denial of Service, Man-in-the-Middle attacks} & \textcolor{black}{Data integrity, Data confidentiality, Authentication} & \textcolor{black}{Privacy} \\ \hline
\textcolor{black}{VANETs} & \textcolor{black}{Communication and Road infrastructure} & \textcolor{black}{Denial of Service, Eavesdropping assault, Node impersonation} & \textcolor{black}{Data Accuracy, Authentication} & \textcolor{black}{Data verification, Privacy, Scalability, Availability} \\ \hline
\textcolor{black}{OBU} & \textcolor{black}{Sharing inter-vehicle safety related information} & \textcolor{black}{Sybil Attack} & \textcolor{black}{Data Accuracy, Authentication} & \textcolor{black}{Access Control} \\ \hline
\textcolor{black}{RSU} & \textcolor{black}{Controlling and supervising the whole network} & \textcolor{black}{Distributed Denial of Service} & \textcolor{black}{Data Accuracy} & \textcolor{black}{Access Control} \\ \hline
\end{tabularx}%
}
\end{table*}

\subsubsection{Privacy}
As CAVs advance, privacy will remain an essential concern \cite{karnouskos2017privacy}. Despite the consistent improvement of self-governing vehicles, it is still unclear what personal data might be gathered via CAVs \cite{taeihagh2019governing}. Different kinds of data that could be gathered via CAVs include the following.

\begin{itemize}
\item \textbf{Owner and passenger information:} This relates to information about the owner and passenger(s). The CAV would need to collect this information regarding authorized users. There are several legislations regarding this aspect, which could apply to certain aspects of CAV data and communications. However, due to the type of data, the manner of collection, or the entity collecting the data for CAVs, these laws may not be applicable.

\item \textbf{Location tracking:} Here, the privacy issue emerges from the ability to view travel data and patterns over time, which can enable a malicious adversary to obtain other information about the vehicle owner or passengers, such as their residence and work as well as other establishments that they frequently visit. Determining the present location or historical travel patterns of a particular person could make them more susceptible to physical harm if the information is accessible to the wrong person. In this aspect, an individual's reasonable expectation of privacy includes concerns about how location data is collected and shared.

\item \textbf{Sensor data:} In AVs, their sensors constantly collect data regarding their environments; Unfortunately, the sensors of AVs can be prone to hacking, leading to potential data leakage. 
\end{itemize}

However, there have been multiple solutions for fortifying the privacy of CAVs that were implemented by researchers with successful results. 

\subsubsection{Data validation}
CAVs generate a lot of data from their sensors that must be collected and analyzed. Furthermore, the data from CAVs need to be verified by the vehicle and/or user that obtained it to prevent any false reporting \cite{cao2020special,yazdinejad2021federated}. CAVs are reliant on their sensors to assess their environment. Unfortunately, this means that the CAVs tend to trust the provided information from the sensors to formulate navigation decisions without data validation \cite{koopman2016challenges}. This leaves the CAV open to attackers; They could feed inaccurate sensor data to disrupt or take control of the system.

Data validation is also an issue in training and testing CAVs \cite{koopman2016challenges}. After all, a CAV is expected to have different behavioral conduct during activity than what was shown during testing \cite{deliparaschos2020preliminary}. Current validation approaches cannot handle this circumstance because they need to consider all conceivable system behaviors in advance in the plan and approval measure for the independent vehicle \cite{shao2019evaluating}. Designers of CAVs would need to test for additional long hours of exposure and gather and validate that much data for validating is a significant issue. The data validation process must also check that the CAVs can recognize nearby or distant objects and guarantee that they will perform effectively in suboptimal climate conditions or degraded environment settings \cite{sheehan2019connected}.
One way to make data validation easier for CAVs is by simulation. Simulation environments comprise several modules (e.g., Environment module, Vehicle module, Physics engine, Sensor module) \cite{gurumurthy2020integrating}. Simulation and modeling tools greatly benefit data validation in the automotive domain due to their high-quality frameworks for design, analysis, data training, and validation \cite{castro2019enabling}. However, simulation has its disadvantages---models trained solely on simulation data fail to generalize to the real world due to the differences between the simulated and realistic environments regarding both physical and visual entities \cite{chao2019force}. 
\begin{figure*}[h]
    \centering
	\includegraphics[width=7in]{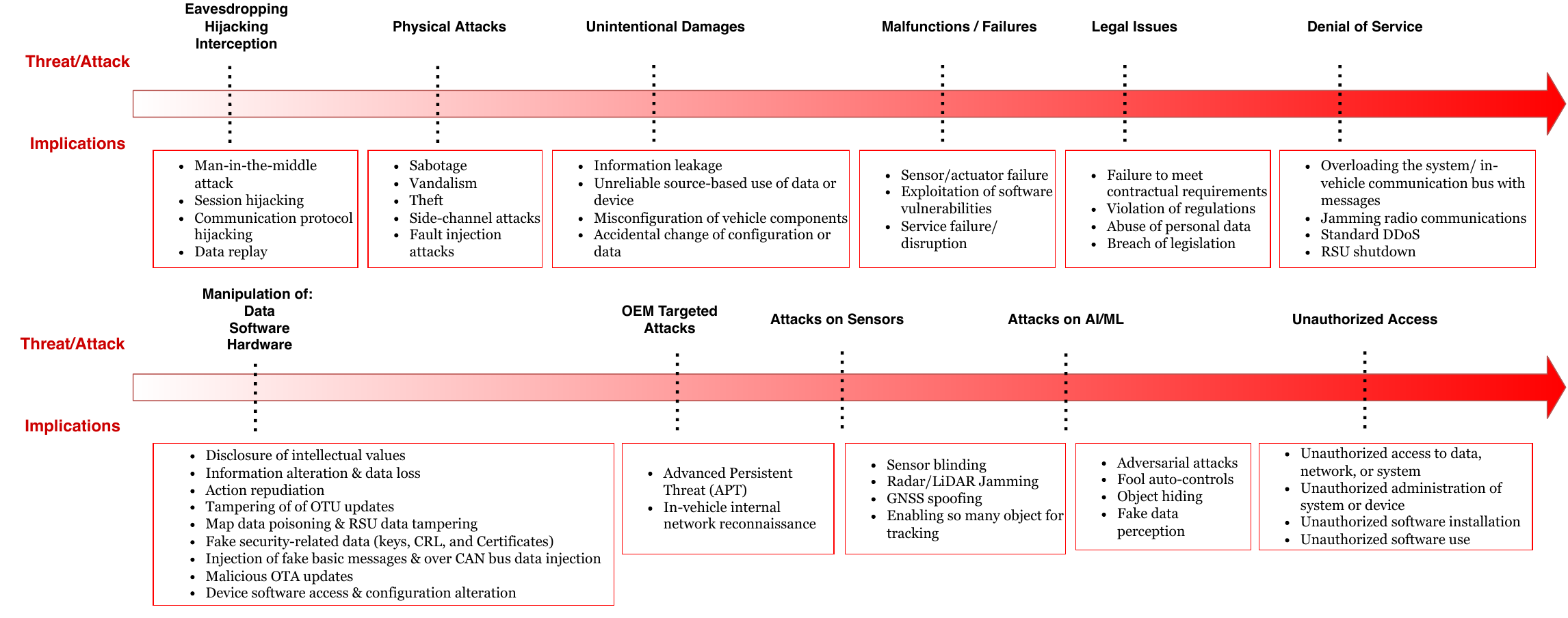}
	\caption{Taxonomy of threats and attacks against CAVs: visualizing the attack landscape.}
	\label{fig:threat-taxonomy}
\end{figure*}

\subsubsection{Liability}
While many works constantly emphasize the need for secure CAVs, these works also gloss over the fact that every decision made regarding the employment of security for CAVs, including the decision to provide little to no security at all, has consequences for the liability of both the manufacturers of CAVs, and those who are involved in implementing the security measures (e.g., algorithms, software, hardware). Security and liability are often correlated, especially concerning automobile accidents. In a typical car accident scenario, liability largely falls on the driver. In CAVs, other entities must be considered, including the manufacturer, software provider(s), service technician(s), and the vehicle owner.
\textcolor{black}{A significant challenge in the context of liability is determining the proportionate responsibility among various stakeholders when a security breach occurs. For instance, in cases where a CAV’s decision-making process leads to an accident, it can be difficult to attribute fault to the vehicle’s AI, its sensors, or the broader communication infrastructure. The introduction of autonomous decision-making further complicates legal responsibility, as traditional concepts of driver liability no longer apply. One potential approach to mitigating liability concerns is the development of transparent and auditable security mechanisms that track decision-making in real-time, allowing for post-incident analysis. Furthermore, regulatory bodies need to evolve with these technological advancements to define clear liability frameworks that consider the distributed nature of CAV systems, ensuring that responsibility is appropriately assigned to manufacturers, software developers, or operators depending on the failure's origin.}

\subsubsection{Malware Detection}
Malicious updates pose a critical danger to CAVs since such malware can cause accidents that may be life-threatening. Furthermore, the network of ECUs for CAVs is indeed linked to both the telematics and infotainer systems to which all the outside devices, vehicles, and infrastructures are connected. This high amount of connectivity gives the attacker multiple paths to have the malware enter the vehicle \cite{kiran2020cyber}. Attackers can also replace legitimate software with malicious software to get control over the CAV \cite{qureshi2021euf}. Malware detection also correlates with the safety of the CAV because attacks look for even the most trivial of bugs to gain access to the CAV. This, in turn, will lead to the endangerment of people's lives - both inside the vehicle and among other people on the road. As a result, security updates that address potential malware issues for CAVs must be robust to properly guarantee the CAVs' safety \cite{9618005}.
\textcolor{black}{One of the key challenges in malware detection for CAVs is the rapid evolution of malware techniques, which can bypass traditional signature-based detection systems. Advanced persistent threats (APTs) that target vulnerabilities in the communication between CAV components—such as the infotainment system and the ECU—require sophisticated, real-time detection mechanisms. Machine learning-based anomaly detection systems show promise in identifying malware behaviors that deviate from established norms. However, these systems must be designed to minimize false positives, which can lead to unnecessary interruptions in the vehicle’s operations. Another critical area of focus is securing the supply chain for software updates, ensuring that every stage—from code development to deployment—incorporates strong integrity checks to prevent the injection of malware into CAV systems. Future research should also explore distributed malware detection techniques that leverage the collaborative nature of CAV ecosystems, where vehicles share information about detected threats in real time, enhancing collective security.}

\subsubsection{Over-the-Air (OTA) Updates}
\textcolor{black}{Communication features enable secure over-the-air (OTA) updates for CAVs. OTA updates utilize communication channels to securely deliver software patches, firmware updates, and security patches to vehicles. Secure communication protocols are employed to ensure the integrity and authenticity of the updates, preventing unauthorized modifications and reducing the risk of tampering or malware injection during the update process. CAVs often receive software updates over the air, which introduces potential vulnerabilities if not properly secured. Malicious updates or compromised update mechanisms could lead to unauthorized access or control of the vehicle's systems \cite{halder2020secure}.}
\textcolor{black}{While OTA updates offer significant convenience and efficiency in maintaining the security and functionality of CAVs, they also introduce several security risks. A major concern is the potential for a man-in-the-middle attack during the update transmission, where an attacker intercepts the update and injects malicious code. To mitigate these risks, strong encryption and multi-factor authentication must be used to verify the authenticity of the update source and protect the integrity of the data during transmission. Another issue is rollback attacks, where an attacker forces the system to revert to an older, vulnerable version of the software. Implementing secure version management and update logs, which are cryptographically protected, can prevent such attacks. Moreover, the reliance on OTA updates emphasizes the need for redundant security measures that ensure a CAV can operate safely even if an update fails or is compromised. Further exploration into blockchain-based secure update mechanisms may offer decentralized, tamper-proof solutions for managing OTA updates in the future.}

\section{Taxonomy of Communication Security Attacks on CAVs} \label{attacks} 
\textcolor{black}{In this section, we present a taxonomy of communication security attacks on CAVs and briefly discuss the various types of attacks. Figure \ref{fig:threat-taxonomy} illustrates this taxonomy.} Generally, several components are involved in an attack on CAVs. The first component is \textit{the attacker}, which is the source of the attack on the CAV. In the event of a system error, it is crucial not only to mitigate and resolve the issue but also to identify the attacker. \textcolor{black}{Identifying the attacker helps reduce the likelihood of future attacks and offers insights into attack vectors (discussed next). Strengthening proactive security measures is therefore critical} \cite{thing2016autonomous,kn}. 

The next component is \textit{the attack vector}. The attack vector represents the path through which an adversary gains unauthorized access to a CAV. This can occur through physical or remote wireless access, enabling exploitation of vulnerabilities within the CAV \cite{ren2019security}. Another important aspect of understanding and addressing security attacks on CAVs is the motivation behind them. According to \cite{pham2020survey}, there are three common motivations for security attacks on CAVs:
\begin{enumerate}
\item \textbf{Interrupting operations:} The goal here is to compromise critical components required for autonomous driving, rendering the autonomous driving mode inoperable.
\item \textbf{Gaining control over CAVs:} In this scenario, attackers seek to gain control of the CAV, allowing them to manipulate its behavior, such as altering the vehicle's route, speed, or triggering emergency brakes.
\item  \textbf{Stealing information:} The adversary aims to collect confidential information from the CAV, which can be used for future attacks. This may also include theft of the user's personal data.
\end{enumerate}

Furthermore, the components of the AV (Autonomous Vehicle) play a significant role in security attacks on CAVs \cite{zang2019impact}. \textcolor{black}{These components are crucial because many of them are vulnerable to attacks} \cite{article88, parkinson2017cyber}. Examples of these components include GPS, Lidar, MMW Radar, Ultrasonic Sensors, Cameras, OBD, ECUs, and CAN systems. Table \ref{tab:AV-Challenge-Security-Components} summarizes the components' capabilities, their tasks, and the types of attacks they are susceptible to. \textcolor{black}{Notably, some components are vulnerable to multiple types of attacks, including spoofing and jamming} \cite{lim2018autonomous}. Additionally, attackers may use certain components, like the ECU, as a means to carry out their attacks \cite{sommer2019survey}. Other critical factors to consider include the importance of assets, the difficulty of attacks, detection rates, and the severity of both information leakage and physical damage, as elaborated in Table \ref{tab:attackseverity}. Table \ref{tab:avattack-taxonomy} provides a summary of various attacks on CAVs.

\textcolor{black}{Moreover, categorizing communication security attacks on CAVs into intra-vehicle and inter-vehicle attacks is challenging because many attacks affect both levels of communication. CAVs rely on complex communication systems that include intra-vehicle communication (within the vehicle) and inter-vehicle communication (between vehicles or with infrastructure) \cite{taslimasa2023security}. Numerous attacks target the communication channels and protocols used by CAVs, aiming to compromise the entire vehicular network, rather than focusing exclusively on intra- or inter-vehicle communication \cite{wang2022security}. For example, an attack on a CAV's internal communication system could hinder its ability to receive critical information from external sources or transmit data to other vehicles or infrastructure. Similarly, an attack on inter-vehicle communication could disrupt the coordination and safety of multiple vehicles within the network. Therefore, communication security attacks on CAVs often transcend the strict boundaries of intra-vehicle and inter-vehicle communication, making it difficult to categorize them strictly within one domain. An in-depth, general taxonomy of these attacks is presented next.}

\begin{table*}[!t]
\centering
\tiny
\caption{CAV attack threats categorized in severity from low (L), medium (M), to high (H).}
\label{tab:attackseverity}
\resizebox{\textwidth}{!}{%
\begin{tabularx}{\textwidth}{|X|X|X|}
\hline
\rowcolor[HTML]{DAE8FC} 
\multicolumn{1}{|l|}{\cellcolor[HTML]{DAE8FC}}    & \multicolumn{1}{c|}{\cellcolor[HTML]{DAE8FC}\textcolor{black}{\textbf{Focus}}} & \textcolor{black}{\textbf{Severity}} \\ \hline
\textcolor{black}{\textbf{Asset Importance}} & Importance of each component & \textcolor{black}{L - AV not affected, M - may influence functions, but no direct impact on operation, H - will cause direct damage to the vehicle} \\ \hline
\textcolor{black}{\textbf{Attack Difficulty}} & Ease of conducting the attack & \textcolor{black}{L - no relevant knowledge needed, not time-consuming, M - time-consuming, but requires a short time to learn required knowledge, H - extensive knowledge required} \\ \hline
\textcolor{black}{\textbf{Detection Rate}} & \textcolor{black}{Odds of detecting attacks via the users or vehicle} & \textcolor{black}{L - no function affected, M - some functions are affected, but the vehicle can detect it and warn users, H - attacks will influence operational functions immediately} \\ \hline
\textcolor{black}{\textbf{Severity of Information Leakage}} & \textcolor{black}{Type of data leaked and how much of the data was leaked} & \textcolor{black}{L - no private information is leaked, M - unimportant information is leaked, H - crucial confidential information is leaked} \\ \hline
\textcolor{black}{\textbf{Severity of Physical Damage}} & \textcolor{black}{Amount of damage sustained from the security attack, if any} & \textcolor{black}{L - not likely to cause physical damage to the user or vehicle, M - small hazards and damage is caused, but not fatal, H - high possibility of causing fatal injuries} \\ \hline
\textcolor{black}{\textbf{Recovery Rate}} & \textcolor{black}{Time needed to recover back to normal after attack is detected.} & \textcolor{black}{L - seconds, M - minutes to hours, H - hours to days} \\ \hline
\end{tabularx}%
}
\end{table*}

\subsection{Zero-day Exploit}
This attack is geared toward potential vulnerabilities. In a Zero-day exploit, vulnerabilities are critical and can result in other devastating security attacks until that vulnerability is taken care of \cite{malik2020analysis}. For hacking CAVs, hackers can use a Zero-Day attack to affect either additional CAVs, the network of CAVs, or other programs, to transfer malware, and to pose serious threats \cite{hou2019zero}. 
\textcolor{black}{A key challenge with Zero-day exploits is that they remain undetected until after an attack occurs, which leaves CAVs vulnerable for extended periods. Additionally, the interconnected nature of CAV networks means that a single exploit can spread quickly across multiple vehicles, amplifying the impact. Effective countermeasures, such as predictive threat intelligence and behavior-based anomaly detection, need to be developed to identify these vulnerabilities before they are exploited. Continuous software updates and patching strategies are essential to mitigate such risks.}


\subsection{Replay Attacks}
Replay attacks send out a previously stored message to deceive other components on the CAV network. Messages used can include elements such as previous updates on traffic or road conditions, accident alerts, etc. \cite{porter2020detecting}. As such, the attack usually stores these types of messages and re-transmits them at different locations or times, so the components believe that these conditions are happening now, even though they occurred in the past. What makes this particular attack dangerous for CAVs is that replay attacks can be used in combination with other attacks, making the impacts more severe \cite{ren2019security}.
\textcolor{black}{To defend against replay attacks, CAV systems need to incorporate time-sensitive authentication mechanisms that ensure messages are only valid for a specific timeframe. Moreover, cryptographic solutions, such as nonces or timestamps, can be implemented to validate the freshness of the transmitted data. Combining these techniques with intrusion detection systems can enhance the CAV network's resilience against such attacks. Real-time monitoring of communication patterns is essential to identify unusual message duplication or delays.}


\subsection{Relay Attacks}
Relay attacks usually target the CAVs' communication systems \cite{ren2019security}. A relay attack is capable of breaking distance restrictions in communication systems. They do so by placing devices between a signal sender and receiver and relaying signals between them. Simply put, an attacker intercepts communication between two entities and, without viewing or manipulating them, relays it to another component in the CAV. This type of attack can also target keyless entry systems of CAVs \cite{ahmad2020securing}.
\textcolor{black}{Relay attacks can be particularly dangerous in keyless entry systems, as attackers can unlock or start vehicles without physical access. To prevent such attacks, manufacturers are implementing technologies like signal jamming or limiting communication range. Additionally, using cryptographic methods to validate the proximity between sender and receiver can mitigate the effectiveness of relay attacks. Enhanced encryption and frequency-hopping techniques are also promising defenses.}


\subsection{DoS Attacks}
This is one of the most common security attacks targeting any layer of the network model \cite{biron2018real}. The idea of a Denial of Service (DoS) attack is to flood the communication network with irrelevant data so that the network cannot respond to genuine users. It is highly adaptive because this type of attack confuses and dodges anomaly detection systems \cite{ali2018intelligent}. DoS uses multiple devices to execute the same attack from different locations. DoS may result in damages to the vehicle and can potentially shut down the entire communication network. Many types of DoS attacks come in various types, such as a flooding attack, jamming attack, and others \cite{cao2020special}.
\textcolor{black}{A significant threat of DoS attacks in CAVs is that they can disrupt critical vehicular functions such as collision avoidance and navigation. This poses serious safety risks, especially in high-speed or congested environments. To counteract these attacks, CAVs need to implement more robust network segmentation, where high-priority messages are isolated from low-priority traffic. Additionally, AI-driven anomaly detection can help identify early signs of DoS attacks by analyzing unusual network behavior.}


\subsection{Sybil Attack}
The Sybil attack is considered one of the most dangerous attacks on CAV networks. This attack involves falsifying the identity of the vehicle/user to disrupt the normal operation of the network \cite{feng2017method}. There are two ways to execute a Sybil attack: one way is by assuming multiple identities to provide false information and have it appear like the information is coming from multiple devices, and the second way is impersonating another's identity and providing false information \cite{lim2020sybil}. Sybil attacks exploit both the topology and mobility of the network and can provide false information about its location; using that, it can provide false information about events at those locations \cite{cao2020special}. 
\textcolor{black}{Further, by creating multiple fake identities or nodes, adversaries in a Sybil attack can deceive the CAV network, leading to several security vulnerabilities \cite{suo2020location}. Firstly, fake identities can disrupt the establishment of secure communication channels between vehicles (inter-vehicular communication). These malicious nodes can inject false messages or interfere with the transmission of legitimate messages, causing confusion, misinformation, or even DoS. Secondly, Sybil attacks can compromise the trustworthiness and integrity of the communication network by spreading false information \cite{bouchouia2023survey}. For instance, fake identities can provide inaccurate data about a vehicle's position, speed, or intentions, leading to potentially dangerous situations such as collisions or incorrect decision-making by other vehicles \cite{bouchouia2023survey}.}


\subsection{Blackhole Attack}
In this scenario, an attacker coerces other components into transmitting as much network traffic as possible to an attacker node \cite{tobin2017approach}. A blackhole attack is executed by informing other entities in the network that the attacker node has the shortest route to their objective. Then, when an entity sends data packets to the attacker, all of them are dropped. This leads to data loss for the network. A blackhole attack causes critical damage by dropping essential data packets, for example, accident alerts \cite{nanda2019internet}.
\textcolor{black}{To mitigate blackhole attacks, CAV networks can use multipath routing techniques, where data packets are transmitted through multiple paths instead of relying on a single route. This reduces the likelihood of data being lost entirely if one path is compromised. Additionally, continuous monitoring of routing behaviors and real-time detection of anomalies in data flow can help identify blackhole nodes early. Implementing reputation-based systems can also prevent malicious nodes from misleading the network by dynamically adjusting trust levels for each node.}


\subsection{Passive Eavesdropping Attack}
This kind of attack is known as a reconnaissance or stealth attack and involves collecting as much information about a network as possible by constantly monitoring the network traffic. The attack analyzes the traffic to recognize patterns and deduce information regarding the vehicles \cite{he2020towards}. This information is then used to carry out attacks by either identifying key entities or replaying recorded network transmissions to act as a smokescreen to avoid detection. \textcolor{black}{Insecure or unencrypted communication channels within the vehicle (intra-vehicle communication) are the key enablers for attackers to eavesdrop on sensitive information. This includes data on passenger privacy, location, or personal preferences \cite{benyahya2022automated}.}
\textcolor{black}{To prevent passive eavesdropping, CAV systems must employ strong encryption protocols for both intra- and inter-vehicular communication. Regular updates to encryption algorithms are critical to staying ahead of evolving eavesdropping techniques. Additionally, implementing secure key exchange protocols will ensure that even if communication is intercepted, the attacker cannot decrypt the data. Lastly, continuous monitoring of network traffic and the use of obfuscation techniques can help obscure patterns that attackers might otherwise exploit.}


\subsection{Side-channel Attacks}
A side-channel attack is a type of security attack that exposes important information in terms of transmitted data or the inner mechanisms of the CAV system via alternative routes \cite{chattopadhyay2017security}. A side-channel attack tries to obtain information indirectly and mainly exploits information leakage \cite{thing2016autonomous}.
\textcolor{black}{Side-channel attacks are particularly challenging because they do not target the software directly but rather exploit physical phenomena such as power consumption, timing, or electromagnetic emissions. To defend against such attacks, CAV systems need to implement hardware-level countermeasures like noise generation and power equalization, which make it difficult for attackers to deduce meaningful information from side-channel data. Additionally, securing cryptographic operations from side-channel leakage is critical in ensuring that sensitive operations cannot be monitored. Future research should focus on developing side-channel attack-resistant architectures for CAVs.}


\subsection{Code Modification and False Data Injection}
These attacks target the OBD-II scanner and affect the ECU \cite{thing2016autonomous}. Specifically, the OBD-II scanner enables personnel to perform diagnoses of CAVs, and more advanced versions can modify a part of the ECU, such as the ECU code. Unfortunately, adversaries can exploit these tools to execute malicious code modifications to compromise the CAV \cite{gupta2015survey}. \textcolor{black}{Further, adversaries may attempt to inject false data into the communication network, misleading vehicles about road conditions, traffic patterns, or obstacles. This can disrupt the cooperative behavior of autonomous vehicles and potentially cause accidents \cite{al2020intelligence}.}
\textcolor{black}{Code modification attacks can be mitigated by securing diagnostic tools like OBD-II scanners with encryption and access control mechanisms, ensuring that only authorized personnel can modify the ECU. Additionally, integrity-checking algorithms should be implemented to detect unauthorized changes to the ECU code in real-time. False data injection attacks can be countered through redundancy in data sources, where vehicles cross-check information from multiple sensors or other vehicles to detect inconsistencies. Securing sensor data with cryptographic signatures can also help prevent tampering.}


\subsection{Packet Sniffing}
In CAVs, particularly for communication, a packet sniffer is an entity that can intercept and log traffic transferred over a communication link. Packet sniffers are typically used for diagnosing network-centric issues \cite{tayeb2017securing}. In simpler terms, packet sniffers allow viewing communication information between two or more nodes. While useful, an attacker can use packet sniffers to access unencrypted data and obtain information. Packet sniffers can also be used for a replay attack \cite{thing2016autonomous}.
\textcolor{black}{To prevent packet sniffing attacks, CAV networks must employ end-to-end encryption for all communication channels, ensuring that even if data is intercepted, it remains unreadable to attackers. Furthermore, intrusion detection systems can be configured to monitor for abnormal data interception activities, triggering alerts when unauthorized sniffing is detected. Employing secure tunneling protocols such as VPNs can further add a layer of protection to sensitive communication. Ensuring timely updates of encryption protocols and implementing secure key exchange methods is also critical.}


\subsection{Sensor Spoofing}
\textcolor{black}{A sensor spoofing attack can severely impact intra-vehicular communication security in a CAV. In such an attack, adversaries manipulate or deceive the sensors within a vehicle to provide false or misleading information \cite{limbasiya2022systematic}. This can compromise the integrity of the intra-vehicular communication system, leading to several security threats. Specifically, sensor spoofing can impact the perception capabilities of the vehicle, causing it to misinterpret the surrounding environment \cite{yang2023anomaly}. For instance, if the spoofed sensors provide inaccurate data about the vehicle's position, speed, or the presence of obstacles, it can lead to incorrect decision-making and potentially hazardous situations. Additionally, the compromised sensors can propagate false information within the vehicle's internal communication network, affecting the cooperation and coordination among different systems and components \cite{pham2021survey}.}

\subsection{Packet Fuzzing}
In a packet fuzzing approach, invalid data gets sent over to the targeted CAV component(s) or module(s) to explore the likelihood of triggering an error condition or fault \cite{tomlinson2018using}. These error conditions result in vulnerabilities that are exploitable and reveal security loopholes. Ironically, Packet fuzzing can also be used for security testing. When used for testing the security of CAVs, fuzzing is implemented for detecting potential problems in the CAV system that could contribute to security issues at a later stage \cite{thing2016autonomous,ff}.

\subsection{Communication Jamming}
\textcolor{black}{Malicious actors can jam communication signals between vehicles or with infrastructure systems, disrupting the flow of critical information and resulting in communication breakdowns \cite{khan2020cyber}. Here, adversaries deliberately interfere with the wireless communication signals between vehicles (inter-vehicular communication), causing disruptions or complete DoS. This type of attack can disrupt the exchange of critical information among vehicles, leading to compromised safety, coordination, and situational awareness. By jamming the communication channels, adversaries can prevent vehicles from receiving important messages, such as collision warnings, traffic updates, or cooperative maneuvering signals. As a result, CAVs may fail to respond appropriately to changing road conditions or the presence of other vehicles, increasing the risk of accidents \cite{pham2021survey}. Moreover, jamming attacks can create confusion within the vehicular network, as vehicles may lose connectivity with each other and the infrastructure, leading to a breakdown in cooperative operations and coordination.}

\subsection{Attacks on Global Positioning Systems (GPS) and Localization}
The GPS delivers position data with an accuracy level. An increased number of satellites has been deployed to provide better GPS coverage and overcome the reported difficulties and obstruction impact in connected vehicles \cite{parkinson2017cyber}. Although the public GPS domain is an openly accessible standard, rogue signals can be feasibly carried out to block or mislead GPS-based devices. This adversarial conduct leads to jamming and spoofing attacks, which involve disseminating wrong yet valid and realistic GPS signals to misguide the receivers. As such, the adversary can broadcast synchronized signals (i.e., synchronization with valid signals observed by the target receiver) to alter the target position increasingly.

Moreover, the creation of plug-and-play attack devices is becoming a reality since the GPS spoofing rewards are augmenting in parallel, and theories on how to conduct the attack are already publicly available \cite{tippenhauer2011requirements}. Examples of proof-of-concept attacks can be concluded from the existing literature, e.g., establishing a counterfeit GPS signal to overpower genuine signals of the GPS incrementally.

Last, anomaly detection for GPS is another solution proposed to enhance autonomous driving security \cite{manale2022intrusion}. The key motivation is to provide sufficient security measures for the GPS sensor, as protecting an environment in constant motion will be challenging in regard to securing against attacks and detecting threats. A model is proposed in \cite{manale2022intrusion} to help secure the GPS sensor based on deep learning. The model further managed to improve both the accuracy and scalability of the vehicle.

\subsection{Adversarial ML Attacks on Perception Systems}
This subsection discusses examples of adversarial ML attacks on CAVs' perception systems. Typically, the adversarial attacks on ML comprise two categories: poisoning attacks and evasion attacks. The poisoning attacks primarily impact the training stage during the learning process. The adversary here aims at compromising the learning process of ML/DL via a successful manipulation of the data used in the training phase \cite{biggio2012poisoning}. Unlike poisoning attacks, evasion attacks mainly target the inference stage of the learning process. The adversary in this scenario aims to create a false output by manipulating the real-time inputs or the test data in the ML model \cite{biggio2013evasion}. Note that the examples utilized in falsifying the ML/DL applications during the inference stage are generally called adversarial examples.

Further, an attacker may aim for a partial perturbation of the images collected by cameras in a CAV. The goal is to make the ML/DL implemented for the vehicle's vision application produce incorrect predictions. Although the DL modules implemented within Electronic Control Units (ECUs) are the primary targets of this adversarial attack, cameras also enable adversarial channels to inject malicious images at ease. Additionally, adversarial patches and stickers with patterns were successful in fooling a broad range of artificial algorithms \cite{liu2018dpatch}. An adversary can easily print and bind these pattern-based adversarial stickers to critical transportation objects like road and traffic signs.

\begin{table*}[!t]
\centering
\tiny
\caption{A taxonomy of existing security attacks against CAVs.}
\label{tab:avattack-taxonomy}
\resizebox{\textwidth}{!}{%
\begin{tabularx}{\textwidth}{|X|X|X|}
\hline
\rowcolor[HTML]{DAE8FC} 
\textcolor{black}{\textbf{Attack}} & \textcolor{black}{\textbf{Main Target(s)}} & \textcolor{black}{\textbf{Impact(s)/Implications}} \\ \hline
\textcolor{black}{Blinding} & Camera & Disruption of visual environment representation \\ \hline
\textcolor{black}{Spoofing} & \begin{tabular}[c]{@{}l@{}}GPS\\Lidar\\MMW Radar\\Ultrasonic Sensors\end{tabular} & \begin{tabular}[c]{@{}l@{}}Disseminate wrong but realistic signals\\Create counterfeit GPS signals\end{tabular} \\ \hline
\textcolor{black}{Jamming} & \begin{tabular}[c]{@{}l@{}}GPS\\MMW Radar\\Ultrasonic Sensors\end{tabular} & \begin{tabular}[c]{@{}l@{}}Disseminate wrong yet realistic signals\\Misguide signal receivers\\Perform plug-and-play attack\\\end{tabular} \\ \hline
\textcolor{black}{Relay} & Lidar & \begin{tabular}[c]{@{}l@{}}Break distance restrictions in systems \\ Target keyless entry systems of vehicles\end{tabular} \\ \hline
\textcolor{black}{Malicious OBD Device} & OBD Port & \begin{tabular}[c]{@{}l@{}}Explicit violation of logical constructs\\and controls to manipulate the vehicle's \end{tabular} \\ \hline
\textcolor{black}{DoS} & \begin{tabular}[c]{@{}l@{}}Intra/inter-communication\\CAN\\OBD ports\end{tabular} & \begin{tabular}[c]{@{}l@{}}Target the CAN protocol\\Transmit volume of high-priority messages\\via OBD ports\\Disrupt processing of legitimate traffic\end{tabular} \\ \hline
\textcolor{black}{MITM} & Intra/inter-communication & \begin{tabular}[c]{@{}l@{}}Access unencrypted data\\Intercept messages exchanged between vehicles\end{tabular} \\ \hline
\textcolor{black}{Remote Exploitation} & OBD Port & Execute code on telematics via long-range wireless connections \\ \hline
\textcolor{black}{Zero-Day Exploit} & Intra-Vehicle Network & \begin{tabular}[c]{@{}l@{}}Transfer malware\\Cause average to critical threats\end{tabular} \\ \hline
\textcolor{black}{Replay} & Intra-Vehicle Network & \begin{tabular}[c]{@{}l@{}}Deceive vehicle's network components\\Perform location-enabled message tampering\end{tabular} \\ \hline
\textcolor{black}{Sybil} & Intra-Vehicle Network & \begin{tabular}[c]{@{}l@{}}Falsify vehicle/user's identity to disrupt\\network operation\\Exploit topology and mobility of the network\\Deliver false information in events and locations\end{tabular} \\ \hline
\textcolor{black}{Blackhole} & Intra-Vehicle Network & \begin{tabular}[c]{@{}l@{}}Coerce components into transmitting packets\\to the attacker node\\Drop essential packets (e.g., accident alerts)\end{tabular} \\ \hline
\textcolor{black}{Eavesdropping} & Intra-Vehicle Network & \begin{tabular}[c]{@{}l@{}}Constant monitoring of network traffic\\Recognize patterns and deduce information\\Replay recorded packets to avoid detection\end{tabular} \\ \hline
\textcolor{black}{Side-Channel} & Intra-Vehicle Network & Expose data and inner mechanisms via alternative routes \\ \hline
\textcolor{black}{Code Modification} & \begin{tabular}[c]{@{}l@{}}OBD-II scanner\\ECU\end{tabular} & Execute malicious modifications of code \\ \hline
\textcolor{black}{Sniffing} & Intra-Vehicle Network & \begin{tabular}[c]{@{}l@{}}Gain access to unencrypted data\\Perform replay attacks\end{tabular} \\ \hline
\textcolor{black}{Packet Fuzzing} & Intra-Vehicle Network & \begin{tabular}[c]{@{}l@{}}Transmit invalid data\\Trigger an error condition or fault\end{tabular} \\ \hline
\textcolor{black}{Adversarial ML: Poisoning \& Evasion} & \begin{tabular}[c]{@{}l@{}}Sensing\\Image Construction\end{tabular} & \begin{tabular}[c]{@{}l@{}}Compromise the image learning process\\Target the inference stage of the image learning\end{tabular} \\ \hline
\textcolor{black}{CAN Access via OBD-II} & CAN & \begin{tabular}[c]{@{}l@{}}Perform DoS attack\\Perform eavesdrop attacks\\Transmit unauthorized data\end{tabular} \\ \hline
\textcolor{black}{ECU Access via CAN} & ECU & \begin{tabular}[c]{@{}l@{}}Compromise ECUs on the CAN\\Perform code injection, code reprogramming\end{tabular} \\ \hline
\textcolor{black}{CAN Access via Telematics ECUs} & CAN & \begin{tabular}[c]{@{}l@{}}Compromise the CAN\\Bypass intrusion detection and cryptography\end{tabular} \\ \hline
\end{tabularx}%
}
\end{table*}

Several research efforts have further demonstrated the feasibility of misleading AI algorithms for traffic sign classification. Notably, unlike the attack setups carried out by authors in \cite{lu2017no}, which were not successful in compromising the stop signs through a neural network (NN) model, authors in \cite{eykholt2018robust} were more successful in their proposed attack mechanism. Specifically, the authors decorated the traffic signs using light black and white adversarial patches, which led to the learning algorithm failing to identify the stop signs by a moving vehicle efficiently.

The authors in \cite{kelarestaghi2019intelligent} have performed a similar adversarial scenario by remotely leveraging the electromagnetic interference while leaving traffic signs physically unmodified (i.e., without adversarial patches). The authors' setups made the adversarial attack easy to conduct but challenging to identify. Although there are little to no experiment explorations against the CAVs, such an adversarial model can be applied to them with ease. This is of great concern in CAVs as they become inefficient in identifying traffic alerts or even falsely read speed limits. Furthermore, adversarial ML-based sensor attacks on the Lidar-based perception in autonomous driving can lead to semantically impactful security results in CAV settings. Here, an attack goal could be fooling the Lidar-enabled perception to perceive untruthful obstacles before a victim's CAV. As a result, the driving decisions will, in turn, be maliciously manipulated. An example of such attack implementation is presented in \cite{cao2019adversarial}. The authors herein conduct the Lidar-based perception system attack by targeting the front-near false obstacles near the front of a victim CAV (i.e., assuming this scenario can provide the highest potential to launch an immediate faulty driving decision). Moreover, Lidar spoofing attacks are considered in the attack model presented in \cite{cao2019adversarial} since they are proven practical attacks against Lidar perception sensors \cite{shin2017illusion}. Lastly, laser aiming can also be performed in this attack scenario. Here, the adversary can deploy techniques such as camera-based object identification and tracking \cite{cao2019adversarial}. From the CAV settings perspective, such adversarial attacks are stealthy as the shooting-based laser devices are significantly small, and laser pulses are considerably invisible.

Currently, various adversarial mechanisms exist to tamper with ML implementations for object detection tasks, like traffic signs. Such mechanisms are usually referred to as universal adversarial perturbations \cite{sitawarin2018darts, kos2018adversarial, moosavi2017universal}. Last, an IBM research group published Adversarial Robustness 360 Toolbox, an open-source library developed in Python to simulate and thwart adversarial attacks on ML-based image tasks \cite{nicolae2018adversarial}. Figure \ref{fig:threat-taxonomy} provides a summary and visual taxonomy of the existing security attacks and threats against CAVs, along with their primary targets, impacts, and security implications. Additionally, when determining the severity of attacks, it is helpful to base these attacks off specific parameters such as those in Table \ref{tab:attackseverity}. Throughout this sub-section, the attack taxonomy and current attacks on CAVs were discussed. While the majority of CAV components are prone to jamming and spoofing, it is essential to distinguish between the types of jamming and spoofing attacks the components of CAVs are prone to. Not every jamming or spoofing attack works the same way when attacking a specific component of a CAV.
\section{Communication Security Solutions and Evaluation Tools} \label{securityChallenge}  
\textcolor{black}{Communication features play a crucial role in security solutions in CAVs \cite{p4}. For instance, secure communication protocols, such as Transport Layer Security (TLS) or Datagram Transport Layer Security (DTLS), establish encrypted and authenticated communication channels between CAVs and infrastructure. These protocols utilize cryptographic algorithms and digital certificates to ensure secure identification, authentication, and key exchange between communicating entities.} However, CAVs face numerous formidable challenges, including liability, privacy preservation, real-time response, data validation, and malware detection, as previously highlighted in Table \ref{tab:AV-Challenge-Security-Components}. Many solutions have been developed in this regard, and our discussion in this section will be focused on these advances.

\subsection{Privacy}

\textcolor{black}{Communication features are employed to ensure secure data transmission and privacy preservation in CAVs \cite{yazdinejad2022block}. Encryption mechanisms, such as symmetric or asymmetric encryption, protect the confidentiality and integrity of data transmitted between vehicles, infrastructure, and other entities. Privacy-enhancing technologies, such as data anonymization or pseudonymization, can also be applied to protect the privacy of individuals while enabling necessary communication and information exchange. An overview of existing solutions aiming at CAV communication privacy enhancement is provided next.}

\subsubsection{Location Privacy- Stealthy Tracking of CAVs}

In \cite{luo2020stealthy}, the authors specifically focus on location privacy. The authors emphasize that the location privacy of CAVs can be compromised via software side-channel attacks. Side-channel attacks happen when the vehicle's driving software and the attack software share a hardware platform and can reveal the physical properties of a vehicle or its environment. Interestingly, this type of attack exploits a vulnerability in the Adaptive Monte-Carlo Localization (AMCL) algorithm, which has been widely utilized in CAVs. To prove their point, the authors simulate a side-channel attack; To evaluate the effectiveness of their attack, the authors used real-world data collected on a Nissan Leaf driving around Oxford. This was to ensure the attack was tested in a realistic environment. Upon testing, the authors achieved a route prediction accuracy of 81\%.
\textcolor{black}{The high accuracy of this attack highlights a critical need for enhanced defenses against location tracking in CAVs. Strengthening isolation between software components or implementing hardware-level countermeasures can reduce the risk of side-channel attacks. Furthermore, future work should explore machine learning techniques that can detect and mitigate side-channel vulnerabilities before they are exploited. Proactive identification of software-level leaks will be essential in preserving CAV location privacy \cite{zz4}.}

\subsubsection{Privacy-Preserving Three-factor Authentication and Key Agreement}

In \cite{jiang2020unified}, the approach to privacy is cloud-based. Specifically, the authors propose a cloud-based architecture incorporating three-factor authentication and a key agreement protocol known as CT-AKA, the proposed framework integrates elements such as passwords, biometrics, and smart cards to ensure authorized access to both the cloud and CAVs. The biometric approaches the authors incorporated include fuzzy vault, fuzzy commitment, and fuzzy extractor to ensure there is no leakage of the users' privacy \cite{yazdinejad2023secure}. Furthermore, the authors' proposed framework includes two session keys: one between the user and CAV to ensure secure remote control of the CAV, and the other is negotiated between a mobile device and the cloud to make sure that security parameters are not compromised and to ensure cloud data access securely with a high guarantee of security. 

The CT-AKA protocol has 6 phases: (1) system setup, (2) CAV registration, (3) user registration, (4) user authentication, (5) password and bio-metrics change, and (6) smart card revocation. In (1), the cloud boots the CAV system and initializes the system parameters. Next, in (2), the cloud generates and distributes a secret key for each CAV. In (3), the cloud issues a smart card storing the secret key to each user. For (4), the three-factor authentication is executed to verify the user identity and build secure channels among the cloud, the CAV, and the user. To verify the effectiveness of the proposed architecture, the authors used a tool called ProVerif, a type of protocol verifier that verifies mutual authentication between users, the cloud, and CAVs, and ensures the secrecy of the session key they have negotiated. Results indicated the proposed architecture does achieve secrecy and mutual authentication between the involved entities. Additional testing also revealed that the authors' proposed architecture is efficient even with a large number of users. 
\textcolor{black}{The use of ProVerif to validate the protocol adds credibility to the framework’s security claims, making it a viable option for real-world applications. However, implementing such a three-factor authentication system at scale might introduce latency concerns, especially for users accessing CAVs in high-demand environments. Future research could focus on optimizing the framework to reduce latency while maintaining strong security guarantees. Additionally, biometric data security must be prioritized to ensure that sensitive personal information is not exposed in the event of a breach \cite{zz5}.}

\subsubsection{SDN-based Secure and Privacy-Preserving Scheme}

In the work of \cite{garg2019sdn}, the authors focus on the privacy of vehicular networks, which are quickly becoming incorporated into CAVs regarding communication. 5G has also been incorporated into CAVs due to its promise to make CAVs faster, smarter, and safer \cite{chen2017vehicle, hashemi2017out}. Unfortunately, 5G still faces challenges in guaranteeing reliable and secure communications for CAVs \cite{yu2016optimal,shah20185g, molina2017lte}. The authors propose a Software-Defined Networks (SDN) centric architecture that provides end-to-end security and privacy in 5G-enabled networks for CAVs. The proposed architecture is meant to simplify network management while also achieving optimized network communications. There are two phases of operation: First, there is a cryptographic-based authentication protocol that authenticates authority in SDN-centric CAV setups. Then, there is an intrusion detection module that is meant to reduce computational complexity and determine possible intrusions in the network. The authors did an extensive evaluation of their framework using three simulators. Metrics that were evaluated included overall security in terms of the detection rate, false positive rate, accuracy, time of detection, and communication overhead. The authors' proposed framework outperforms other frameworks (in terms of higher performance and lower communication overhead), making this framework highly applicable for securing privacy for CAVs.
\textcolor{black}{The integration of SDN into 5G-enabled networks offers substantial benefits for network management and privacy. However, one of the key challenges is the centralized nature of SDN, which could introduce a single point of failure in the system. Future research should explore decentralized SDN solutions to mitigate this risk. Additionally, optimizing the intrusion detection module to handle emerging 5G-based threats will be critical for maintaining robust security in CAV networks \cite{zz6}.}

\textcolor{black}{With such an SDN-assisted security solution, communication features can play a role in the following contexts:
\begin{itemize}
    \item \textbf{SDN}: Allowing for dynamic, centralized management of network traffic, enhancing security by quickly adapting to threats and potentially isolating malicious nodes thanks to its decentralized and decoupled infrastructure.
    \item \textbf{5G}: Offering faster speeds, lower latency, and better support for device density than its predecessors. For vehicular networks, this translates to quick and reliable communication between vehicles (V2V), between vehicles and infrastructure (V2I), and with other networks.
    \item \textbf{Dynamic configuration}: With SDN's centralized control capabilities, the network can be dynamically reconfigured in response to observed traffic patterns, potential security threats, or specific vehicle requirements, thus enhancing the overall security and performance of vehicular networks.
    \item \textbf{Quality of Service (QoS) management}: Given the critical nature of some vehicular communications, SDN can prioritize certain types of messages, ensuring timely and reliable delivery, especially in scenarios where latency or bandwidth is a concern.
\end{itemize}}

\subsection{Data Validation}

\subsubsection{SAVIOR} The authors of \cite{quinonez2020savior} propose an architecture called SAVIOR to address data validation. SAVIOR is focused on both securing CAVs and data validation. The SAVIOR framework is inspired by what is known as the Physics-based Attack Detection (PBAD) model, where PBAD has two major steps. The first step is performed offline and extracts the physical components of the CAV to generate a model that captures expected correlations between the inputs and the outputs of the system. The second step is online and involves an anomaly detection algorithm that compares prediction with observed states and raises an alarm when the discrepancy between the predicted and the observed states exceeds a given threshold. What is most impressive about this framework is that it applies to both aerial and grounded CAVs. In the offline stage of this proposed architecture, parameters of the physical components of the vehicle are learned. The online stage is where the model is used to predict sensor measurements and compare them accordingly. Next, a prediction is made via the EKF and compared to the observed state. Lastly, the anomaly detection test shows if the differences between the observed and expected data are significant over time.

\textcolor{black}{Robust physical invariants suggest using consistent physical or behavioral properties of vehicles to ensure their security. Communication features can play a role in the following contexts:
\begin{itemize}
    \item \textbf{Data exchange and verification}: Communication protocols can be established where vehicles frequently exchange data about their physical state or surrounding environment. Using robust physical invariants, each vehicle can verify the authenticity of the received data, ensuring it matches expected physical behaviors or constraints.
    \item \textbf{Secure V2V and V2I}: In a CAV environment, secure V2V and V2I communications are vital. They can be designed to utilize physical invariants as a layer of verification, ensuring that received messages are consistent with the known physics of vehicles.
    \item \textbf{Anomaly detection}: By continuously monitoring and comparing communicated data against established physical invariants, it's possible to detect anomalies that might indicate malicious activity or sensor spoofing.
    \item \textbf{Tamper-proof logging}: Communication features can be used to maintain tamper-proof logs based on physical invariants, ensuring a secure record of vehicle behaviors and states for post-analysis or forensic purposes.
    \item \textbf{Feedback and control}: Secure communication channels might provide real-time feedback to the vehicle's control system if deviations from expected physical invariants are detected, enabling corrective actions or emergency measures.
    \item \textbf{Collaborative verification}: In dense CAV environments, neighboring vehicles can collaborate through secure communication channels to verify each other's reported physical states, providing an additional layer of security through distributed consensus.
\end{itemize}}

\subsubsection{Validation Database - Relevant Traffic Scenarios}

In \cite{putz2017system}, P{\"u}tz \textit{et al.} tackles the challenge of data validation by adopting a database-centric approach. Specifically, the authors implement a database called PEGASUS, which addresses the challenge of collecting and validating data for CAVs. According to the authors, The PEGASUS database contains relevant traffic scenarios for training CAVs. PEGASUS combines different database entities and a data processing approach that determines the required test specifications based on various types of collected input data. The types of input data collected could include traffic simulation data, driver simulation data, field data, and others. 

In step 1, the data owner generates a common environment and traffic description. Only the data owner can convert the recorded data into a suitable format. Step 2 is the first step for data processing and validation. It is concerned with the correct data formatting and indexes and assigns access rights based on data provider requirements. Step 3 involves enriching the raw data to interpret potential scenarios CAVs may encounter. Step 4 involves calculating the likelihood of a situation, and Step 5 aids in scenario interpretation. Step 6 is where the scenarios are clustered into predefined logical scenarios, and the parameters of the recorded scenario events get transferred into frequency distributions. Step 7 is where the testing and data validation happens.

\subsubsection{Simulation-Based Platform - Algorithm Validation}

While not explicitly focused on the `data' validation aspect, the work \cite{banerjee2019development} explores another aspect of data validation that the previous examples did not cover: the algorithms used for data validation. The algorithms used for validating data of CAVs are crucial to ensure the data validation process operates as smoothly as possible, and the algorithms themselves are efficient and flexible on the type of data formats used for CAVs. The author notes that developing algorithms for data validation of CAVs requires large quantities of training data. Still, the cost of collecting such training data limits the size and diversity of this data. To remedy this, the author implements a simulation-based platform for testing and validation that combines the autonomous driving simulator (CARLA)  and their own approach. The proposed platform manages to be capable of both validation and development of data validation and autonomous driving algorithms.

\subsection{Liability}
\subsubsection{Liability Attrition Model}
In \cite{oham2018blockchain}, liability for CAVs is discussed and categorized into two main types:
\begin{itemize}
\item \textbf{Product liability:} This refers to damages caused by product defects such as design and manufacturing shortcomings. In this case, the auto manufacturer is liable. If the car accident occurs while the vehicle is in autonomous mode, the software provider is liable, especially when a software program is determined to have led to the accident.
\item \textbf{Negligence liability:} In this case, this refers to damages due to neglecting to execute an action. Here, the vehicle owner is liable when they do not execute an instruction from an auto manufacturer or software provider. 
\end{itemize}

Furthermore, the authors also design and propose a liability attribution model. From there, the authors then implement a blockchain approach for liability attribution. In doing so, the authors' proposed architecture is resilient to evidence tampering, false information, and unavailable evidence. In doing so, this proposed framework fortifies security for CAVs by providing untampered evidence and allows for multiple participants to agree on evidence needed to process insurance claims. This architecture could be applicable in digital forensics, especially since liability for incidents that involve CAVs is a currently developing area of law.

\subsection{Malware Detection}
\subsubsection{eUF} In \cite{qureshi2021euf}, the authors develop a framework called Enhanced Uptane Framework (eUF) for the detection of malicious software updates in CAVs. The proposed solution uses a combination of transfer learning and Convolutional Neural Networks (CNNs) to enhance security. Regarding its architecture, there are two ECUs in eUF, the primary and the secondary. The primary ECU contains more storage and computational resources, so most functions are performed on this part. Meanwhile, the secondary ECU performs fewer functions (e.g., verification of hashes) to reduce the risk of man-in-the-middle attacks inside the CAV. Regarding the CNNs the authors used, the authors trained both CNNs for 20 epochs. Additionally, for training and testing of eUF, the authors created two datasets by collecting executables of Windows and Linux operating systems. Overall, the proposed framework can distinguish between malicious and benign software executables with high accuracy.

\textcolor{black}{Given such a solution, the communication features play a role in the following key contexts:
\begin{itemize}
    \item \textbf{Authentication and verification}: A secure communication protocol would be established to authenticate OTA update sources. Cryptographic signatures might be used to ensure the update's legitimacy.
    \item \textbf{Encrypted communications}: To protect the updated data during transmission, encryption can be used to safeguard against eavesdropping or tampering.
    \item \textbf{Integrity checks}: After an OTA update is received, integrity checks can be performed to confirm the data hasn't been altered during transmission.
    \item \textbf{Behavior analysis}: Post-update behavior of the vehicle can be monitored to detect anomalies or deviations from expected behaviors. If the updated software behaves suspiciously, it might indicate a malicious update.
    \item \textbf{Version control and rollback}: Maintaining a secure record of software versions can help roll back to a previous, trusted state if a malicious update is detected.
    \item \textbf{Notification and alerts}: If a potential malicious OTA update is detected, it could have mechanisms to alert the vehicle operator, manufacturer, or other relevant entities.
    \item \textbf{Collaborative detection}: Such a solution could employ information from a fleet of vehicles. If multiple vehicles experience issues post-update, this collective information can aid in faster detection of malicious updates.
\end{itemize}}

\subsubsection{Security Architecture for Protecting CAVs from Malware} In \cite{8746516}, the authors propose a characterization of CAV malware and security architecture to protect the CAV from various malware. The authors attempt to characterize CAV malware in terms of intention, propagation, exploitation, types, and methods of attacks. Furthermore, the proposed architecture uses multiple computational platforms and the virtualization technique to limit the attack surface. There is also a real-time operating system (OS) to control critical vehicle functionalities and multiple other operating systems for non-critical functionalities (infotainment, telematics, etc.). The security architecture also describes groups of components for the operating systems to prevent malicious activities and perform policing (monitor, detect, and control).

\subsection{Performance Evaluation Tools for Secure Vehicle Communications}
Here, we discuss some evaluation tools that have also been developed for CAVs. Having these evaluation tools would provide further insight into the operation of the CAV, as well as its strengths and shortcomings regarding security. Interestingly, there do not seem to be too many evaluation tools that have been proposed. As such, in the future, more attention needs to be paid to designing evaluation tools for CAVs. Evaluation tools also need to account for current methods used in communication. Furthermore, when designing evaluation tools, they need to be multi-purpose, as some only have a few purposes.

\subsubsection{SEPAD}
Proposed in \cite{zelle2020sepad}, this specific system is interesting since it is less of an `architecture' but more of a platform for assessing security for self-sufficient driving. An interesting point when planning security systems for self-governing vehicles is what new technologies can the autonomous systems support \cite{rosique2019systematic}. Further, one also has to consider the alteration of parts of the self-sufficient vehicle, which is done either by coordinating new security arrangements or by performing assaults. However, there is a risk of damage to the vehicle. This is why many researchers implement theoretical models when developing security frameworks \cite{jing2020agent}. Unfortunately, the drawback of using theoretical models is that the results and data obtained may not mirror real CAVs \cite{chao2020survey}. So the creators proposed the Security Evaluation Platform for Autonomous Driving (SEPAD) to help scientists demonstrate their security models. Hence, newly created security components can be effortlessly incorporated and assessed in a sensible situation without danger to the specialist. Additionally, the architectural design choices can be demonstrated and evaluated in terms of security, leading to more precise assessments and helping researchers gain a better understanding of how to incorporate security into autonomous vehicles.

SEPAD uses numerous communication technologies as well, such as the IEEE 802.11p network to handle vehicle-to-vehicle communication. Authentication is covered by using ISO 15118 communication technology. SEPAD also contains 5 modules that can be used to enhance a secure development process, resulting in stronger and more intuitive mechanisms. These modules are: Simulate Attacks, Analyze Impact, Evaluate Risk, Adapt Security Strategy, and Evaluate in Operation; The simulated attacks module uses an attack simulation component of SEPAD to mimic relevant attacks from the attacker. Then, the Analyze Impact module simply analyzes the impact of the attack. The Evaluate Risk component deals with risk for the vehicle, the passengers, external persons, and the overall environment. It also provides information about the probability of an attack by providing data about the preconditions for an attack or how easy it is to execute in realistic environments. Then based on whatever risk(s) are found, the adapt security strategy module can be used to modify current security mechanisms accordingly. In the Evaluate in Operation module, an evaluation of the operating car is performed without implementing attacks to identify any possible issues introduced by the security mechanisms. Then the process repeats itself.

It is of utmost importance that one can evaluate the performance of a CAV's ability to secure communications \cite{huang2016autonomous, d2018analytical}. However, there does not seem to be much literature on performance evaluation tools used for evaluating a CAV's performance in securing their communication. This can present problems in verifying whether the CAVs are operating correctly and whether communication in CAVs is secure (and any possible data leakage is minimized). The following subsections address the topic of developing performance evaluation tools for securing CAV communication in detail.

\subsubsection{V2X Testing}
The authors of \cite{wang2019survey} specifically focus on V2X application technology for CAVs. V2X employs current information and communication technology to enable V2V, vehicle to V2I, vehicle to pedestrian (V2P), and vehicle to network/cloud (V2N/V2C) network connections \cite{molina2017lte, chen2017vehicle}. Specifically, the authors emphasize that the V2X technology is rigorously tested. A notable testament to the article is that it focuses on current evaluation methods for V2X technologies for CAVs. An issue that the authors noticed is that the current evaluation approaches and tools used are independent, and each of them has only one or two testing purposes. The authors propose an end-to-end evaluation tool that combines virtual and real environments that can handle the testing task of the full communication protocol stack. While the article effectively discusses current evaluation tools for V2X testing, it falls short of detail regarding their proposed evaluation tool. 

\subsubsection{BEHAVE}
While the research work in \cite{ivanchev2019need} does not explicitly focus on communication (the article concerns traffic analysis), it does provide an architecture for an evaluation tool called BEHAVE. Technically, BEHAVE is a simulator, but simulation tools have been used in the performance evaluation of CAVs. Therefore, this article is applicable to be included in this section. Interestingly, the authors note that available models and evaluation tools are not sufficient to fully evaluate mixed traffic scenarios because most of them focus only on the lower level control of the CAV \cite{ghiasi2019mixed}. Securing CAVs' communication is crucial for mixed traffic analysis because the CAV needs to communicate traffic data to its sensors, such as the changing of red light to green light, the location of other CAVs about traffic lanes, speed limit, and other factors \cite{bhavsar2017risk}. The BEHAVE evaluation tool is powered via the agent-based traffic simulator CityMoS. Additionally, the BEHAVE evaluation tool also has a data monitoring component that accounts for different traffic scenarios. The framework also includes a way to visualize and evaluate the evolution of agent-specific parameters for traffic analysis.

\subsubsection{VeRA}
Risk analysis was studied as an essential use case related to developing CAVs \cite{9140383}. Risk analysis/assessment concerns itself with determining whether the risk of an attack on the CAV is critical or minor. The main argument is that current risk analysis methods are either too time-consuming or not suitable for CAVs. To remedy these issues, a security risk analysis method called Vehicles Risk Analysis (VeRA) is developed to help evaluate the risks of attacks in the context of CAVs. VeRA uses a simpler analysis process and fewer factors in comparison to other benchmarks (e.g., SAE J3061), significantly reducing required analysis time without negatively impacting accuracy. A mathematical model is also deployed in this use case to assess risk value whereas the previous methods used the tedious process of looking up tables. VeRA can obtain the same analysis results in 43\% less time compared to three other risk analysis methods-EVTA, RACE, and CSRL.

\subsection{Analysis of Security Frameworks for CAVs}
As we conclude our discussion on communication security solutions for CAVs, it is essential to evaluate how the proposed solutions compare with existing frameworks. The comparative analysis Table \ref{tt} provides a side-by-side comparison, highlighting the distinct advantages and potential limitations of each approach. This table synthesizes the key attributes of each framework, offering a comprehensive overview that aids in understanding their applicability in addressing the specific security challenges faced by CAVs. Such analysis is crucial for informing future developments and guiding the selection of appropriate security measures tailored to the needs of the CAV ecosystem
\begin{table}[htbp]
\small
\centering
\caption{Comparative Analysis of Security Frameworks for CAVs}
\scalebox{0.8}{
\label{tab:comparison}
\begin{tabular}{@{}|p{2cm}|p{3cm}|p{3cm}|p{3.5cm}|p{3.5cm}|@{}}
\hline
\rowcolor[HTML]{DAE8FC} \multicolumn{5}{|c|}{\textbf{Security Frameworks Comparison}} \\ \hline
\textbf{Feature} & \textbf{Proposed Solutions} & \textbf{Existing Solutions} & \textbf{Benefits} & \textbf{Drawbacks} \\
\hline
\textbf{Scalability} & Highly scalable using distributed architecture & Often centralized, less scalable & Can handle large fleets of CAVs efficiently & More complex to set up initially \\
\hline
\textbf{Robustness} & Utilizes advanced encryption and AI-based anomaly detection & Relies on traditional encryption methods & Higher security against modern cyber threats & Potentially higher computational overhead \\
\hline
\textbf{Flexibility} & Modular design allows for easy updates & Rigid frameworks with infrequent updates & Quick adaptation to emerging threats & Requires continuous development effort \\
\hline
\textbf{Ease of Implementation} & Requires initial setup and training of AI models & Generally simpler to deploy & Customizable to specific fleet needs & Steeper learning curve for system operators \\
\hline
\textbf{Addressing Vulnerabilities} & Specifically targets intra- and inter-vehicular communications & Broad focus, not specific to CAVs & Enhanced protection for CAV-specific threats & May overlook broader cybersecurity concerns \\
\hline
\end{tabular}}\label{tt}
\end{table}


\section{Communication Security Protocols for CAVs and Best Practices} \label{protocols}
This section discusses prominent security protocols that have been newly developed for CAV and summarizes best security design practices.

\subsection{Security Protocols}
The typical well-known protocols include (but are not limited to): CAN (Controller-Area Network), LIN (Local Interconnect Network), FlexRay, MOST (Media Oriented Systems Transport), DSRC (Dedicated Short Range Communications), LTE-V (long-term evolution-vehicle), DSR (Dynamic Source Routing), AODV (Ad hoc on Demand Distance Vector), DSDV (Destination Sequenced Distance vector), and GPSR (Geographic Perimeter Stateless Routing). Since these well-known protocols have already been comprehensively discussed in previous work, we discuss new protocols. Table \ref{tab:protocols} gives a brief overview of the well-known protocols and their advantages and drawbacks and Table \ref{tab:proposed-protocols-av} discusses some newly proposed protocols that have been proposed for CAVs.

\begin{table*}[!t]
\centering
\tiny
\caption{Summary of security issues related to well-known vehicular protocols along with their pros and cons.}
\label{tab:protocols}
\resizebox{\textwidth}{!}{%
\begin{tabularx}{\textwidth}{|>{\raggedright\arraybackslash}X|>{\raggedright\arraybackslash}X|>{\raggedright\arraybackslash}X|>{\raggedright\arraybackslash}X|>{\raggedright\arraybackslash}X|}
\hline
\rowcolor[HTML]{DAE8FC} 
\multicolumn{5}{|c|}{\textcolor{black}{\textbf{Intra-Vehicle Communication Protocols}}} \\ \hline
\textcolor{black}{\textbf{Protocol}} & \textcolor{black}{\textbf{Area(s) of Focus}} & \textcolor{black}{\textbf{Advantages}} & \textcolor{black}{\textbf{Drawbacks}} & \textcolor{black}{\textbf{Security Issue(s)}} \\ \hline
\textcolor{black}{CAN} & Performance & Low cost, high reliability, relatively low bandwidth, easy configuration & Lack of message authentication, unencrypted traffic & Denial of Service \\ \hline
\textcolor{black}{LIN} & Remote application & Open source, cheaper than CAN, low complexity, low cost & Low bandwidth & False data injection \\ \hline
\textcolor{black}{FlexRay} & Fault tolerance, communication & High-speed communication, versatile system layouts, error detection capability & More expensive, higher complexity, lower voltage operating levels & Spoofing, Denial of Service, lacks confidentiality and authentication, eavesdropping, masquerading, injection, and replay attacks \\ \hline
\textcolor{black}{MOST} & Vehicular infotainment systems & High speed, flexibility & Limited work regarding security of MOST & Jamming, vulnerable to attacks targeting synchronization \\ \hline
\rowcolor[HTML]{DAE8FC} 
\multicolumn{5}{|c|}{\textcolor{black}{\textbf{Inter-Vehicle Communication Protocols}}} \\ \hline
\textcolor{black}{\textbf{Protocol}} & \textcolor{black}{\textbf{Area(s) of Focus}} & \textcolor{black}{\textbf{Advantages}} & \textcolor{black}{\textbf{Drawbacks}} & \textcolor{black}{\textbf{Security Issue(s)}} \\ \hline
\textcolor{black}{DSRC} & Security, communication & Flexibility, low latency & Low scalability & Jamming, spoofing, interference \\ \hline
\textcolor{black}{LTE-V} & Reliability & Wider coverage, better resource allocation & Unpredictable delays & Message flooding \\ \hline
\rowcolor[HTML]{DAE8FC} 
\multicolumn{5}{|c|}{\textcolor{black}{\textbf{Ad-Hoc Routing Protocols}}} \\ \hline
\textcolor{black}{\textbf{Protocol}} & \textcolor{black}{\textbf{Area(s) of Focus}} & \textcolor{black}{\textbf{Advantages}} & \textcolor{black}{\textbf{Drawbacks}} & \textcolor{black}{\textbf{Security Issue(s)}} \\ \hline
\textcolor{black}{DSR} & Route calculation, route maintenance, route discovery & Less latency, lower delay, lower overhead, efficient, high throughput, lower packet loss & Scalability & Malicious forwarding nodes \\ \hline
\textcolor{black}{DSDV} & Consistency, delivery time, packet rate & Less modification, congestion control & Low packet delivery rate, high overhead & Malicious routing updates, routing table poisoning \\ \hline
\textcolor{black}{GPSR} & Wireless routing & Little memory needed, scalability & Instability, inaccuracy & Location misreporting, excessive packet overload \\ \hline
\textcolor{black}{AODV} & Routing & Better communication performance, reduced communication overhead, quick response to changes & Congestion, instability & Blackhole attacks \\ \hline
\end{tabularx}%
}
\end{table*}


\begin{table}[htbp]
\tiny
\centering
\caption{A review of recent CAV protocols.}
\scalebox{0.8}{
\label{tab:proposed-protocols-av}
\begin{tabularx}{\textwidth}{|>{\raggedright\arraybackslash}X|>{\centering\arraybackslash}X|>{\raggedright\arraybackslash}X|>{\raggedright\arraybackslash}X|}
\hline
\rowcolor[HTML]{DAE8FC}
\multicolumn{4}{|c|}{\textcolor{black}{\textbf{Proposed Protocols}}} \\ \hline
\textcolor{black}{\textbf{Protocol}} & \textcolor{black}{\textbf{Year}} & \textcolor{black}{\textbf{Area(s) of Focus}} & \textcolor{black}{\textbf{Benefit(s)}} \\ \hline
\textcolor{black}{ASC} & 2017 & Authentication & Reduced communication and computational cost, anonymity, lower packet loss ratio \\ \hline
\textcolor{black}{RHR} & 2019 & Network bandwidth & Reduced communication overhead, improved performance, real-time feedback \\ \hline
\textcolor{black}{SAP-IoV} & 2019 & Privacy, Authentication & Better security and performance, reduced time cost \\ \hline
\textcolor{black}{\begin{tabular}[c]{@{}c@{}}Distributed Multi-Channel \\ MAC Protocol for VANET\end{tabular}} & 2019 & Flexibility, Network topology, Control channel access & No excessive overhead \\ \hline
\textcolor{black}{CoMACAV} & 2019 & Data transmission & Increased throughput \\ \hline
\textcolor{black}{NC-MAC} & 2020 & \begin{tabular}[c]{@{}l@{}}V2V/X broadcasting, Beacon transmission \\ and reliability for VANETS\end{tabular} & Improved reliability \\ \hline
\textcolor{black}{MD-AODV} & 2020 & Communication & Lower packet delay \\ \hline
\textcolor{black}{Li-Net} & 2020 & Connectivity & Reduced message delay \\ \hline
\textcolor{black}{CVIP} & 2020 & Scalability & High flexibility \\ \hline
\textcolor{black}{DTMR} & 2021 & Communication Overhead & Stability, applicable in highways and urban environments \\ \hline
\textcolor{black}{PARRoT} & 2021 & Mobility, Predictive Routing & Lower end-to-end latency \\ \hline
\textcolor{black}{TLBGR} & 2021 & \begin{tabular}[c]{@{}l@{}}Data Acquisition, Packet loss, \\ Real-time data transmission\end{tabular} & Avoid data congestion, high delivery rate of data packets \\ \hline
\textcolor{black}{AnonSURP} & 2022 & Location Privacy, Data Privacy & Reduced communication overhead, resistance from various known attacks \\ \hline
\end{tabularx}}
\end{table}

Some of the newer protocols were meant to alleviate the shortcomings of some more well-known protocols. These examples come from \cite{9249520}, \cite{8815628}, \cite{9351930}, \cite{9304556}. These authors proposed newer protocols to fix some shortcomings of the AODV protocols regarding its performance in packet delivery, packet delay, and congestion.

\subsubsection{Li-Fi Vehicle Network (Li-Net)}
This protocol was proposed by \cite{panhwar2020li} and uses Li-Fi, which is expected to have better connectivity, security, and bandwidth in comparison to WiFi. Li-Net is used for V2V and V2I communication. The protocol also tackles various highway scenarios that could occur with CAVs, with a focus on traffic congestion. As such, the protocol has potential use in managing traffic. Li-Net also manages to achieve low packet loss. Notably, Li-Net was tested as a use-case for smart cities in terms of collision avoidance. To further enhance Li-Net, the authors in \cite {janjua2021li} proposed a scheduling method to better handle message transmissions. This proposed method was the Intel Quartus Prime II, which provided a simulation environment for the Li-Fi. Codes were separated for the sender and receiver nodes prior to analyzing their characteristics. The authors in \cite{janjua2021li} also conducted experiments on a regular VANET simulation that was prone to various message delays and message throughput issues. Test results revealed that the proposed scheduling method helped reduce packet loss ratios and average packet delay even more.

\subsubsection{Authentication Based on Smart Card (ASC)}
The authors in \cite{8016361} proposed the ASC protocol, an anonymous and lightweight authentication based on the smart card. A low-cost cryptographic mechanism is used by ASC to authenticate the legitimacy of the users (vehicles) and to validate data messages. ASC consists of five phases: (1) user registration, (2) user login phase, (3) user authentication phase, (4) password change phase, and (5) data authentication phase. According to the authors, the user login and user authentication phases are intended to check for legitimacy. The password change phase aims to resist an offline password-guessing attack. Finally, messages are authenticated among vehicles in the data authentication phase. For testing, the authors used VanetMobiSim and they were able to reduce communication and computational cost by more than 50\%. Further, ASC has a login identity that is changed dynamically to prevent an attack from linking a target vehicle, allowing the protocol to be anonymous. Although the authors provided formal proof of how their protocol treats the security issue, the proof would be dis-proven by the authors in \cite{8603720}.

\subsubsection{Secure Authentication Protocol for Internet of Vehicles (IoV)}
The authors in \cite{8603720} tested the ASC protocol and found that the ASC protocol suffered from numerous attacks (e.g., offline identity guessing attack, location spoofing attack, and replay attack) despite the authors in \cite{8016361} claiming the opposite. The authors in \cite{8603720} also found that the ASC protocol took a considerable amount of time for authentication and had issues with scaling. To fix these issues, a new protocol was proposed the Secure Authentication Protocol for Internet of Vehicles (referred to as SAP-IoV in our paper).  SAP-IoV was supported with formal proof to demonstrate its security. For testing, the authors used an iPhone 6S smartphone as the testing platform; Specifically, the system specifications were: iOS 10.11, CPU Apple A9+M9 co-processor up to 2.1 GHz, RAM 2GB. Overall testing results showed the proposed protocol was more secure and can reduce the authentication time cost.

\subsubsection{Decentralized Privacy-preserving Routing Service Protocol (DPPRSP)}
In \cite{tsao2022private}, the authors propose a decentralized privacy-preserving routing service protocol (referred to as DPPRSP) with a focus on location. The authors emphasize that in most current routing services, users give their individual location data directly to routing services in exchange for route recommendations. Since this data is associated with the users' identity, schedules, habits, preferences and other private information can be inferred through repeated interactions with routing services. This would support the need for privacy preservation so that user's data is not compromised. To provide strong security, DPPRSP uses the Laplace mechanism in combination with Secure Multi-Party Computation (SMPC) to provide privacy and ensure the protocol is cryptographically secure. Although the Laplace mechanism is used for enforcing differential privacy, it tends to return semantically impossible values, such as negative count \cite{phan2017adaptive}. Nevertheless, DPPRSP can obtain the aggregate effect of traffic on travel time, allowing an estimate of travel time accurately and privately. The authors conducted experiments using the Sioux Fall transportation network and were able to demonstrate that DPPRSP manages to provide minimal performance overhead and guarantees privacy. 

\subsubsection{B-IoMV}
The work in \cite{gupta2022b} is unique in the sense the article covers the Internet of Military Vehicles (IoMVs). While this is technically out of the scope of this paper, the article is justified in being included since it addresses relevant concepts, concerns, and methodologies for CAVs. Furthermore, \cite{gupta2022b} incorporates blockchain technology for building the proposed protocol. Specifically, the proposed protocol, B-IoMV addresses issues relating to security and anonymity. Upon testing, the authors of the proposed protocol showed that the B-IoMV system achieved better communication latency, data storage cost, and network bandwidth utilization. 

\subsubsection{AnonSURP}
The authors in \cite{shariq2022anonsurp} propose an anonymous and secure ultra-lightweight RFID protocol for the Internet of Vehicles called AnonSURP. AnonSURP aims to ensure the user's location privacy, and data privacy, and to resist typical known attacks. Upon conducting an informal analysis, the authors were able to confirm that AnonSURP meets the various security requirements and provides resistance to various known attacks. The authors also conducted a comparative analysis against other protocols and noted that AnonSURP outperformed other protocols in terms of performance. 

\subsubsection{Hybrid Software-Defined Networking-based Geo-graphical Routing Protocol (HSDN-GRA)}
The authors in \cite{article0000} propose a novel protocol for CAVs called HSDN-GRA (Hybrid Software-Defined Networking-based Geo-graphical Routing Protocol). The authors note that SDN is optimal for CAVs due to reduced interference, the ability to increase exploitation of wireless channels, and enhanced resource management for networks. HSDN-GRA uses a clustering approach and considers three factors: (1) contact duration between vehicles, (2) each CAV's available load, and (3) log of communication errors. By considering multiple criteria, HSDN-GRA enables choosing the most trustworthy CAVs by avoiding communication problems and ensuring there is connection availability. To test HSDN-GRA, the authors compared it against a few other protocols: MA-DSDV \cite{harrabi2014multi} and PSO-C-MA-DSDV \cite{harrabi2016novel}. Simulation of their network was done on the JADE simulation platform. HSDN-GRA managed to outperform the other two protocols and achieve better throughput and stability and minimal overhead.

\subsection{Lessons Learned}
When designing security protocols for CAV communication, it usually includes the accompanying crucial steps: (1) identify security objectives and requirements of the system; (2) assess the CAV's value and sensitivity to be protected; (3) define security policies in accordance with the security requirements; (4) estimate the potential adversaries capabilities; (5) design those protocols that align with the sensitivity of the system and the risks it is exposed to \cite{chattopadhyay2020autonomous}. Further, when studying the cybersecurity of CAVs, it is useful to recognize steps and cycles where individuals are engaged while considering the CAV's life-cycle security \cite{vskorput2020cybersecurity}. The people, social, and environmental factors are some of the key sources of weaknesses inside CAVs, which can fluctuate from system to system \cite{el2020cybersecurity}.

\subsubsection{Defining Security Objectives}
To improve intra- and inter-vehicular communications, security objectives must consider the management and people aspects for CAVs \cite{chuprov2020reputation}, \cite{ito2020supporting}. As such, it is imperative that when implementing the security protocols, security objectives are designed in a way where everyone can understand what is being required (security is meaningful only when people understand what needs to be focused on \cite{qayyum2020securing}). Having clearly defined security objectives for these protocols not only allows us to consider the cost of potential cybersecurity attacks but also the cost of implementing security mechanisms and the opportunity cost of limiting certain features for protocols \cite{kukkala2020sedan}. Even so, the primary objective for designing security protocols must be concerned with the resilience of the CAVs' safety feature functions \cite{chattopadhyay2020autonomous}.  

\subsubsection{Establishing a Trust Model}
The Trust Model is essential to all useful security designs, including protocols, since it sets up the foundation of who is reliable, permitting clients to figure out what can be trusted, and empowering the architects to figure out who is the trusted authority for executing these protocols \cite{jakaria2020formal}. Security protocols should begin from a reasonable and practical trust model, with cryptographic systems being established afterward to protect the communication of data exchange and guarantee that no malignant assailants can bypass them \cite{halder2020secure}. The trust model is particularly significant in the CAV's communication connectivity infrastructure \cite{dirsehan2020examination, guo2020towards} because the communication connectivity aspect of CAVs includes establishing and preserving identity and privilege management of the involved devices for CAVs \cite{javaid2020scalable}. 

\subsubsection{Establishing a Security Perimeter}
By using a security perimeter, the CAV is segregated into different security domains, each having different threat environments \cite{chowdhury2017secure}. This would improve the design and implementation of security protocols for CAVs' communications in terms of adaptability and flexibility. By establishing security perimeter(s), the proposed security protocols can distinguish between cybersecurity incidents and physical tampering attacks of CAVs \cite{dimase2020holistic}. Ultimately, including security perimeters in security protocols enables a systematic and comprehensible approach to communication security analysis and the overall design of the CAV.

\subsubsection{Establishing a Secure and Trustworthy Network Management Infrastructure}
Security protocols must have a secure and trustworthy network management infrastructure to ensure a balance between preventing malicious attackers from exploiting potential vulnerabilities and efficient communication so that data can be exchanged quickly, allowing enhanced communications \cite{nanda2019internet}. Network management is crucial for security protocols as they must account for resource constraints in CAVs' devices. The resource constraints in the CAV's devices imply that there needs to be accountability for how much bandwidth can be allocated, which security protocols must consider \cite{ge2019ultra}. Although there are several components that need to be included here, we precisely cover the following two, as they are relevant to security and communication protocols.

\begin{itemize}
    \item \textbf{Infrastructure for trust management}: Some primary examples of communication are V2V and V2I \cite{nanda2019internet}. In these types of communications, control systems for admitting new nodes and detecting malevolent nodes are important prerequisites to keep the security protocols intact \cite{liu2020reinforcement}. 
    \item \textbf{Heterogeneous network integration}: Because CAVs are integrated with a variety of devices, they usually have a mix of wireless communication systems, and these often come with their own security protocols \cite{sharma2020heterogeneous}. Hence, when designing security protocols for CAVs, they need to be generalized enough to account for all involved devices. As such, interoperability is an important aspect of security protocols as well since the communication of data for CAVs may require different conversion of data formats \cite{gautham2020heterogeneous,shrestha2020evolution}. 
\end{itemize}

Other best practices should be considered as well, such as communication overhead, throughput, and run-time. These additional metrics should be considered when designing security protocols as a protocol that is secure but slow will result in massive delays in data transmission. While there may not be a guarantee that the protocol will run as fast as possible on every component of the CAV (due to the components' restraints), there should still be effort dedicated to minimizing delays in communication.


\section{Open Issues and Future Directions}


\begin {figure*} [htbp]
\centering
\includegraphics [width=.85\textwidth] {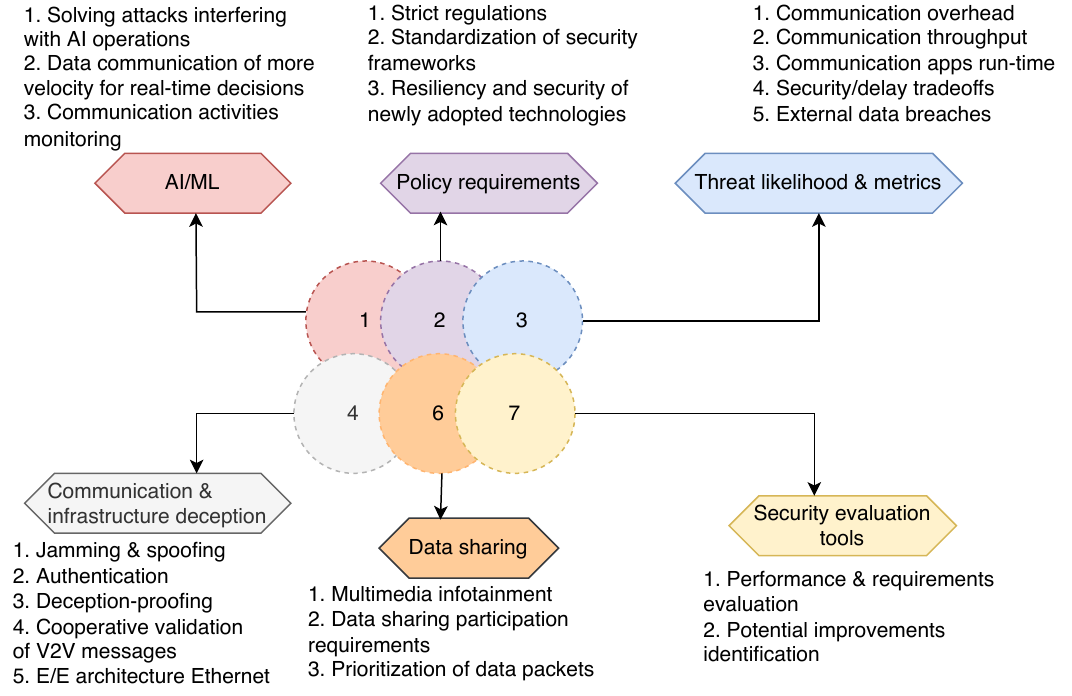}
\caption {Categorizing future research avenues in CAV security: a taxonomy.}
\label{future-dir}
\end {figure*}

Overall, many security architectures have been proposed for CAVs. Unfortunately, there is a lack of standardization for security frameworks for CAVs. This could be because there are multiple aspects to consider when designing security architectures for CAVs, such as privacy, data leakage, communication ranges of CAV devices, and others. As seen in the discussed security architectures, it is imperative to evaluate proposed security frameworks to ensure the framework can be applied to CAVs. New technologies can be incorporated for building security frameworks for CAVs, such as blockchain and edge computing. However these new technologies still need to be flexible, efficient, and secure to ensure less likelihood of data leakage.

Various security challenges have been addressed through existing CAVs' solutions and protocols. However, some challenges remain significant in some of these solutions. The biggest challenges mainly include privacy, real-time response, data validation, and liability. Very few articles have attempted to mitigate these four key challenges with their own implementations, achieving optimal results in the process. A few limitations also persist in some existing vehicular protocols such as the FlexRay protocol. Precisely, there are lower operating voltage levels, which contribute to problems for extending network length. Additionally, some CAV protocols have only been implemented in high-end vehicles. As a result, it is difficult, or even impossible, to predict how the protocol would work for CAVs that are not premium \cite{pullen2019security}.

It is important to implement CAVs' protocols to satisfy higher security and performance requirement levels, such as fault tolerance, dependability, bandwidth, and other related factors for intra- and inter-vehicular communications. The key open challenges and research directions related to CAVs and their communication security are depicted in Figure \ref{future-dir} and discussed in detail next.

\subsection{Strict Regulations}
\textcolor{black}{Unlike traditional vehicles, where the vehicle's key control belongs to the driver, the driving task of CAVs is partially or completely controlled by the autonomous driving system. In order for this system to enable control and decision-making tasks, it significantly depends on the CAV's status (e.g., high precision maps, road conditions, etc.) that is gathered by communication components and sensors in real-time. As a result, a second delay may occur, leading to catastrophic incidents/accidents. Therefore, rigorous requirements and regulations must be designed and implemented to enhance the resiliency and security of CAV communication by minimizing this type of delay within the autonomous driving process.}

\subsection{Standardization of Security Frameworks}
Despite the existence of remarkable security architectures for CAVs, a lack of standardization remains considerable for some of the proposed security frameworks of CAVs as there are various aspects to consider when designing its security architectures, including privacy, data leakage, and vehicular communication ranges. Further, newer technologies can be incorporated in the future to design security frameworks for CAVs, such as blockchain and edge computing. Several existing solutions have already incorporated such technologies. However, future work must consider flexibility, efficiency, and reliability to ensure less likelihood of data leakage.

\subsection{Threat Likelihood and Metrics}
Several existing solutions and protocols are considered reasonably robust against the likelihood of vehicular cyberattacks. Additionally, there are remarkable aspects to be considered in future security protocols and solutions for CAVs. Such aspects mainly include the communication overhead, throughput, and application run-time. By considering these additional metrics, delays associated with data transmission (i.e., end-to-end inter-communication delay) can be lowered. Furthermore, since connectivity is a primary component of CAVs, it is also crucial to ensure the most efficient security architectures are being implemented in a way that communication is secure without adding latency and increasing communication overhead. The security design of CAVs requires that both the individual components and communication technologies used are rigorously protected and that access from outside of the vehicle, if applicable, is strictly controlled. Without proper security architectures, even the strongest communication technologies will be vulnerable to data breaches.

\subsection{E/E Architecture Security}
The current E/E architectures with a large number of ECUs and a single central vehicle gateway are expanding and yet becoming complex. Dedicated domain controllers have been leveraged for these architectures to consolidate ECUs in the area of ADAS and Cockpit Electronics. Yet, future works must consider the automotive Ethernet in this vehicular architecture in terms of wiring harness, complexity, and security.

\subsection{Authentication}
Authentication is another considerable concern to be addressed in the future within wireless communication in CAVs as adversaries can compromise users' data (through false identities). Moreover, communication solutions for CAVs must address the computing power aspect, which is substantial for improving the power consumption and the individual devices involved in the communication of vehicles \cite{s6}.

\subsection{Communication Jamming and Spoofing}
According to the existing attack surface against CAVs, the majority of CAV components are vulnerable to jamming and spoofing. It is important to distinguish between the types of jamming and spoofing attacks the components of CAVs are vulnerable to. Many existing protocols do not consider that in some scenarios, jamming or spoofing attacks do not work the same way when attacking a specific CAV's component. Further, when determining the severity of attacks, many existing security frameworks and vehicular protocols only base these attacks on certain parameters. Hence, further efforts are still needed to secure vehicular communications against jamming and spoofing attacks.

\subsection{Data Sharing and Prioritization}
\textcolor{black}{Multimedia infotainment in traditional vehicle networks mainly relies on service providers, whereas in CAVs, it relies on individual autonomous vehicles sharing their information. Therefore, future works in CAV need to consider the popularity of CAVs, specifying the minimum number of vehicles required to participate in information sharing in order to ensure more real-time data and enable further precise decision-making. The key insight upon this open challenge is that the data sharing rate will remain low if CAVs are not widespread (i.e., a limited number of CAVs). This challenge impacts not only real-time-based decision-making in the autonomous driving task but also its overall communications.}

\textcolor{black}{Currently, the communication channel bandwidth remains relatively low, and the distribution of CAVs is decreasing drastically, potentially leading to a large increase in the number of propagated data packets. This challenge will become even more significant as the distance among CAVs continues to shrink. Hence, the priority of transmitted/propagated data packets must be accurately considered by future works in CAVs to guarantee that vehicles are enabled to have timely access to information that is important to them. Specifically, future priority-based communication solutions may support varying message priorities to include, for instance, periodically broadcast data (e.g., location data), warning data (e.g., dangerous road conditions), and emergency data (e.g., emergency responders vehicle approaching).}

\subsection{Trusted Communications and Deception-Proofing}
\textcolor{black}{It remains unclear when and which communication messages must be trusted in CAVs. CAVs may receive messages related to false incidents and can thus react in a manner endangering other vehicles. Therefore, CAVs must consider their context or situation when replying to other data messages. Although some recent works in the literature attempted to identify and handle false/fake deceptive messages (i.e., verification and validation of V2V communication messages), the context of cooperation has not been considered. Future works in CAV deception-proofing must consider designing cooperative mechanisms that are secure and resilient to invalid messages and misinterpretation of messages received from other vehicles.}

\subsection{Security Evaluation Tools}
Furthermore, additional attention must be given to designing performance evaluation tools for evaluating the CAVs' ability to secure communications. The lack of literature addressing this aspect indicates that another challenge is evaluating how well these security architectures perform apart from designing effective security architectures. Potentially, this could be another area of research regarding the implementation of security for CAVs. Without effective tools to evaluate proposed frameworks, there is no way of knowing where improvements to the proposed frameworks could be made.

\section{Future Roadmap} \label{future} 
As we navigate the evolving landscape of CAVs, we must chart a future roadmap that addresses emerging challenges and harnesses new opportunities in communication security. This section outlines key areas that constitute promising avenues for future research and development:
\begin{itemize}

 \item \textbf{Enhanced Security Protocols}: Future studies should focus on developing more resilient security protocols that adapt to evolving cyber threats. This includes exploring advanced cryptographic methods and AI-based anomaly detection systems to safeguard CAV communications.

 \item \textbf{Integration of Blockchain Technology}: Investigating the integration of blockchain technology for secure and transparent CAV operations. Blockchain can offer decentralized security solutions, ensuring data integrity and trust in vehicle-to-vehicle and vehicle-to-infrastructure communications.

 \item \textbf{Quantum Computing and Post-Quantum Cryptography}: With the advent of quantum computing, current cryptographic standards may become vulnerable \cite{s4}. Future research should explore post-quantum cryptography to prepare CAV systems for quantum-era threats.

 \item \textbf{Autonomous Vehicle Ethics and Legislation}: As technology advances, the ethical implications and legal frameworks governing autonomous vehicles need further exploration. Future research should address the development of comprehensive guidelines and standards that ensure the responsible and ethical use of CAVs.

 \item \textbf{5G and Beyond Networking Technologies}: Investigating how emerging networking technologies, particularly 5G and beyond, can enhance CAV communication systems. This includes research on minimizing latency, improving bandwidth, and ensuring consistent connectivity for seamless CAV operation.

 \item \textbf{Human-Machine Interface (HMI) and User Experience:} Studying the evolving role of HMI in CAVs and its impact on user experience and safety \cite{s5}. This includes user-centric designs, intuitive interfaces, and seamless integration of autonomous functionalities.

 \item \textbf{Cross-Disciplinary Collaboration}: Encouraging collaborative research between computer science, automotive engineering, and cyber-physical systems to foster innovative solutions in CAV security.

 \item \textbf{Simulation and Real-World Testing}: Developing advanced simulation models and conducting extensive real-world testing to validate the effectiveness and reliability of proposed CAV security solutions in various scenarios.

 \item \textbf{Adapting to Environmental Challenges}: Researching the impact of different environmental conditions on CAV communication systems and developing adaptive technologies to maintain consistent performance under varied conditions.
   
\end{itemize}
By addressing these areas, we can propel the CAV industry toward a future that not only embraces technological advancements but also prioritizes security, safety, and ethical considerations.

\section{Conclusions} \label{conclusion} 

\textcolor{black}{This article has presented and discussed the latest developments in security protocols, architectures, and frameworks, highlighting best practices for designing robust security measures for CAVs. As the shift towards CAV adoption accelerates, this paper contributes significantly by categorizing various security frameworks, aiding in the development of secure and intelligent transportation systems.} We have paid particular attention to vehicular communication security, \textcolor{black}{outlining} the challenges and opportunities presented by state-of-the-art solutions in the context of autonomous driving technologies. Specifically, this research has examined both intra- and inter-vehicular communication security challenges, proposing techniques to secure these critical aspects. Our findings enabled us to propose comprehensive security frameworks and outline relevant use cases, \textcolor{black}{offering a foundation upon which future research can expand}. \textcolor{black}{Although considerable progress has been made in enhancing CAV security,} the journey is far from complete. Ensuring the security of CAVs is a dynamic process that requires continuous technical assurances, especially as CAV technologies evolve and new cyber threats emerge. This necessitates that security frameworks be designed with adaptability and scalability in mind, \textcolor{black}{capable of responding to evolving cyber threats}. To advance this field, future research should focus on developing adaptive security algorithms that can dynamically adjust to new threats as they are identified. \textcolor{black}{Furthermore, the integration of artificial intelligence and machine learning techniques has the potential to enhance the ability to predict and mitigate potential attacks before they cause harm.} Collaborative efforts between academia, industry, and regulatory bodies are also essential to develop standardized security protocols that address both current and future cybersecurity challenges in CAV systems. \textcolor{black}{This paper emphasizes the importance of proactive approaches in securing CAVs, setting the stage for ongoing research and innovation in this critical area of transportation technology. The safety and reliability of future transportation depend not only on advancements in vehicle automation but also on the continuous improvement of cybersecurity measures.}



\bibliographystyle{elsarticle-num-names} 
\bibliography{cas}





\end{document}